%% file: main.tex
\documentclass[12pt,a4paper,final]{article}
\pdfoutput=1

\usepackage{amsmath}
\usepackage{amssymb}
\usepackage{amsthm}
\usepackage{exscale}
\usepackage[mathscr]{eucal}
\usepackage[normalem]{ulem}
\usepackage{floatrow}
\usepackage{bm}
\usepackage{graphicx}
\usepackage{color}
\usepackage{dsfont} 
\usepackage{lineno}
\usepackage[font={small}]{caption}
\usepackage[bookmarks=false]{hyperref}

\usepackage{geometry}
\newcommand{\ps}{p}
\newcommand{\Ps}{P_{\mathrm{sub}}}
 
\newcommand{\ga}{\gamma}
\newcommand{\pd}{p_{\mathrm{dis}}}
\newcommand{\Xs}{X_{\mathrm{sub}}}
\newcommand{\Gs}{G_{\mathrm{sub}}}
\newcommand{\gs}{g_{\mathrm{sub}}}
\newcommand{\pb}{p_{\mathrm{bump}}}
\newcommand{\pconn}{p_{\mathrm{conn}}}
\newcommand{\str}{s_{\mathrm{trans}}}

\newcommand{\lams}{\lambda_{\mathrm{sub}}}
\newcommand{\smax}{s_{\mathrm{max}}}

\title{Subsampling scaling: a theory about inference from partly observed systems}
\author{A. Levina$^{1,3*}$ and V. Priesemann$^{2,3*}$
	\\ {\small $^1$Institute of Science and Technology Austria,  Am Campus 1, 3400  Klosterneuburg} 
	 \\ {\small $^2$Max Planck Institute for Dynamics and Self-Organization and}
	 \\ {\small $^3$Bernstein Center for Computational Neuroscience, Am Fassberg 17, 37077 G\"ottingen, Germany}
	\\ {\small $^*$anna/viola@nld.ds.mpg.de; the authors contributed equally}}

\begin{document}
\maketitle

\input{Levina_Priesemann_0927}

\input{SI}

	\begin{small}
			\bibliography{soc_bib}
	\bibliographystyle{unsrt}

\end{small}

\end{document}

%% file: Levina_Priesemann_0927.tex
\section*{Abstract}

In real-world applications, observations are often constrained to a small fraction of a system. 
Such spatial subsampling can be caused by the inaccessibility
or the sheer size of the system, and cannot be overcome by longer sampling.
Spatial subsampling can strongly bias inferences about a system's aggregated properties.
To overcome the bias, we derive analytically a subsampling scaling framework that is applicable to different observables, including distributions of neuronal avalanches, of number of people infected during an epidemic outbreak, and of node degrees.
We demonstrate how to infer the correct distributions of the underlying full system, 
how to apply it to distinguish critical from subcritical systems, and
how to disentangle subsampling and finite size effects.
Lastly, we apply subsampling scaling to neuronal avalanche models and to recordings from developing neural networks. We show that only mature, but not young networks follow  power-law scaling, indicating self-organization to criticality during development.

\section{Introduction}

Inferring global properties of a system from observations is a challenge, even if one can observe the whole system. The same task becomes even more challenging if one can only sample  a small number of  units at a time (spatial subsampling). For example, when recording spiking activity from a brain area with current technology, only a very small fraction of all neurons can be accessed with millisecond precision. To still infer global properties, it is necessary to extrapolate from this small sampled fraction to the full system. 

Spatial subsampling affects inferences not only in neuroscience, but in many different systems: In disease outbreaks, typically a fraction of cases remains unreported, hindering a correct inference about the true disease impact~\cite{Papoz1996, Cormack1999}. Likewise, in gene regulatory networks, typically a fraction of connections  remains unknown. 
Similarly, when evaluating  social networks, the data sets are often so large that because of computational constraints only a subset is stored and analyzed.
Obviously, subsampling does not affect our inferences about properties of a single observed unit, such as the firing rate of a neuron. 
However, we are often confronted with strong biases when assessing aggregated properties, such as distributions of node degrees, or the number of events in a time window ~\cite{Stumpf2005, Priesemann2009, Ribeiro2010,Gerhard2011}. 
Concrete examples are distributions of the number of diseased people in an outbreak, the size of an avalanche in critical systems, the number of synchronously active neurons, or the number of connections of a node. Despite the clear difference between these observables, the mathematical structure of the subsampling problem is the same. Hence our novel inference approach applies to all of them.

Examples of subsampling biases, some of them dramatic, have already been demonstrated in numerical studies. For example, subsampling of avalanches in a critical model can make a simple monotonic distribution appear multi-modal~\cite{Priesemann2009}. In general, subsampling has been shown to affect avalanche distributions in various ways, which can make a critical system appear sub- or supercritical~\cite{Ribeiro2010, Priesemann2013, Priesemann2014,  Ribeiro2014, Yu2014, Wilting2016}, and sampling from a locally connected network can make the network appear ``small-world''~\cite{Gerhard2011}. For the topology of networks,  it has been derived that, contrary to common intuition, a subsample from a scale-free network is not itself scale-free~\cite{Stumpf2005}.  Importantly, these biases are not due to limited statistics (which could be overcome by collecting more data, e.g. acquiring longer recordings, or more independent subsamples of a system), but genuinely originates from observing a small fraction of the system, and then making inferences including unobserved parts. 
Although subsampling effects are known, in the literature there is so far no general analytical understanding of how to overcome them. 
For subsampling effects on degree distributions, Stumpf and colleagues provided a first analytical framework, stating the problem of subsampling bias~\cite{Stumpf2005}.  

In this paper, we show how to overcome subsampling effects. To this end we develop a mathematical theory that allows to understand and revert them in a general manner. We validate the analytical approach using various simulated models, and finally apply it to infer distributions of avalanches in developing neural networks that are heavily subsampled due to experimental constraints. Finally, we show that finite-size and subsampling effects clearly differ, and derived a combined subsampling-finite-size scaling relation. Together, our results introduce a novel approach to study under-observed systems.

\section{Results}


\subsection{Mathematical subsampling}

To derive how spatial subsampling affects a probability distribution of observables, we define a minimal model of ``mathematical subsampling''. 
{   
We first introduce the variables with the example of avalanches, which are defined as cascades of activity propagating on a network~\cite{Beggs2003,Bak1987}, and then present the mathematical definition. 
The main object of interest is a ``cluster'', e.g. an avalanche. The cluster size $s$ is the total number of events or spikes. 
In general, the cluster size is described by a discrete, non-negative random  variable $X$. Let $X$ be distributed according to a probability distribution $P(X=s)=P(s)$.} 
For subsampling, we assume for each cluster that each of its events is independently observed with probability $\ps$ (or missed with probability $1-\ps$). Then $\Xs$ is a random variable denoting the number of \textit{observed} events of a cluster, and $X-\Xs$ the number of missed events.
{   For neural avalanches, this subsampling is approximated by sampling a random fraction of all neurons. Then $\Xs$  represents the number of all events generated by the \textit{observed} neurons within one avalanche on the full system. 
Note, that this definition translates one cluster in the full system to exactly one cluster under subsampling (potentially of size zero; 
  {this definition does not require explicit binning, see Sec.~\ref{sec:binning} and Sec.~\ref{sec:avdef}}). 
We call the probability distribution of $\Xs$ ``subsampled distribution'' $\Ps(s)$. 
An analogous treatment can be applied to e.g. graphs. There a ``cluster'' represents the set of (directed) connections of a specific node, and thus $X$ is the degree of that node. Under subsampling, i.e. considering a random subnetwork, only connections between \textit{observed} nodes are taken into account, resulting in the subsampled degree $\Xs$.}

\begin{figure}
	\begin{center}
		\includegraphics[width=0.95\textwidth]{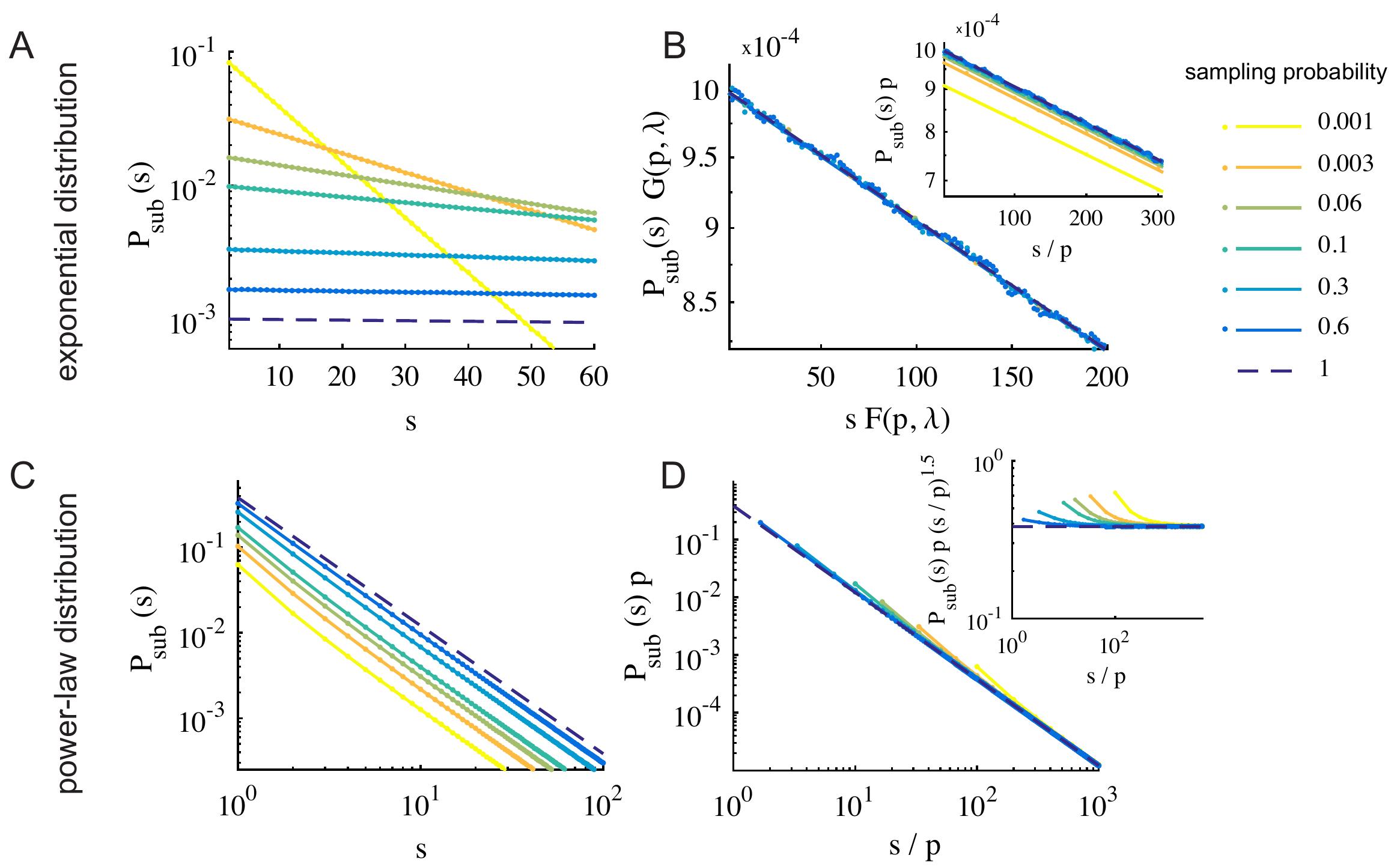}
	\end{center}
	{\caption{Mathematical subsampling of exponential and power-law distributions.
			\textbf{A}: Subsamplings of an exponential distribution with exponent $\lambda=0.001$.  
			\textbf{B}: Collapse of subsampled exponential distributions by subsampling scaling derived in Eq.~\ref{Eq:ExpScal}. Inset: same with p-scaling (Eq.~\ref{Eq:SubsScal}). 			
			\textbf{C}: Subsampled power-law distributions with exponent $\gamma=1.5$. 
			\textbf{D}: Collapse of the same distributions  by p-scaling (Eq.~\ref{Eq:SubsScal}); inset: flattened version.
			Note the log-linear axes in A,B, and the double-logarithmic axes in C,D. 
			{Solid lines are analytical results (Eq.~\ref{Eq:BinSub}), dots are numerical results from subsampling $10^7$ avalanches (realizations of the random variable $X$) of the corresponding original distribution. Colors indicate the sampling probability $\ps$.}
			\label{Fig:MathSub}}}
	
\end{figure}

As each event is observed independently, the probability of $\Xs=s$  is the sum over probabilities of observing clusters of $X=s+k$ events, where $k$ denotes the missed events and $s$ the sampled ones (binomial sampling):

\begin{equation}
P_{\mathrm{sub}}(s)=P(\Xs=s)=\sum_{k=0}^\infty P(s+k) {s+k \choose s}\ps^s (1-\ps)^k. \label{Eq:BinSub}
\end{equation}

This equation holds for any discrete $P(s)$   {defined on $\mathbb{N}_0$, the set of non-negative integers}. To infer $P(s)$ from $\Ps(s)$, we develop in the following a novel ``subsampling scaling'' that allows to parcel out the changes in $P(s)$ originating from spatial subsampling. A correct scaling ansatz collapses the $\Ps(s)$ for any sampling probability $\ps$.

In the following, we focus on subsampling from two specific families of distributions that are of particular importance in the context of neuroscience, namely exponential distributions $P(s) = C_\lambda e^{-\lambda s} $ with $\lambda>0$, and 
 power laws $P(s) = C_\gamma s^{-\gamma}$ with $\gamma>1$. These two families are known to show different behaviors under subsampling~\cite{Stumpf2005}: 

\begin{enumerate}
	\item For exponential distributions, $P(s)$ and $\Ps(s)$ belong to the same class of distributions, only their parameters change under subsampling. Notably, this result generalizes to positive and negative binomial distributions, which include Poisson distributions.
	\item Power-laws or scale-free distributions, despite their name, are not invariant under subsampling. Namely,  if $P(s)$ follows a power-law distribution, then $\Ps(s)$ is not a power law but only approaching it in the limit of large cluster size ($s \rightarrow \infty$).  
\end{enumerate}

In more detail, for exponential distributions, $P(s) = C_\lambda e^{-\lambda s} $, $s \in \mathbb{N}_0$, subsampling with probability $\ps$ results in an exponential distribution with decay parameter $\lams$ that can be expressed as a function of $\lambda$ and $\ps$ (for the full analytical derivation see Sec.~\ref{sec:SupplExp}):

\begin{equation}
\lams=\ln\left( \frac{e^\lambda+\ps-1}{\ps}\right) \Leftrightarrow \lambda =  \ln((e^{\lams}-1)\ps +1).
\label{Eq:lams}
\end{equation} 

Likewise, changes in the normalizing constant $C_{\lambda}=1-e^{-\lambda}$ of $P(s)$ are given by:

\begin{equation}
C_\lambda/C_{\lams}= 1-e^{-\lambda}+\ps e^{-\lambda}=\frac{e^{-\lams}+\ps-\ps e^{-\lams}}{\ps}.
\end{equation}

These two relations allow to derive explicitly a subsampling scaling for exponentials, i.e. the relation between $P(s)$ and $\Ps(s)$:

\begin{align}
P(s)&= \frac{C_\lambda}{C_{\lams}} \Ps\left(\frac{\lambda}{\lams}s\right)= \frac{e^{-\lams}+\ps-\ps e^{-\lams}}{\ps} \Ps\left(\frac{  \ln\left(e^{\lams}\ps-\ps +1\right)}{\lams}s\right) \label{Eq:ExpScal}\\
&= \left(1-e^{-\lambda}+\ps e^{-\lambda}\right)\Ps\left(\frac{ \lambda}{ \ln\left( \frac{e^\lambda+\ps-1}{\ps}\right)}s\right) = G(\ps,\lambda) \Ps(s F(\ps,\lambda)).\nonumber
\end{align}
Thus given an exponential distribution $P(s)$ of the full system, all distributions under subsampling can be derived. Vice versa, given the observed subsampled distribution $\Ps(s)$, the full distribution can be analytically derived if the sampling probability $\ps$ is known.
Therefore, for exponentials, the scaling ansatz above allows to collapse all distributions obtained under subsampling with any $\ps$ (Fig.~\ref{Fig:MathSub}~A,B).

{   The presented formalism is analogous to the one proposed by Stumpf et al.~\cite{Stumpf2005}. They studied which distributions changed and which preserved their classes under subsampling. In the following we extend that study, and then develop a formalism that allows to extrapolate the original distribution from the subsampling, also in the case where an exact solution is not possible.}


For power-law distributions of $X$, $\Xs$ is not power-law distributed, but only approaches a power law in the tail ($s \to \infty$). An approximate scaling relation, however, collapses the tails of distributions as follows (mathematical derivation in Sec.~\ref{sec:SupplPowerLaw}). For $s\to \infty$,  a power law $P(s)=C_\gamma s^{-\gamma}$ and the distributions obtained under subsampling can be collapsed by:
\begin{equation}
P(s)=\ps^a \Ps(\ps^b s), \: \mathrm{~for~any~} a, b \in \mathds{R} \: \mathrm{with} \: a-b \gamma =1-\gamma.
\label{Eq:ab}
\end{equation}
  {For any $a,b$ satisfying the relation above, this scaling collapses the tails of  power-law distributions.} 
The ``heads'', however, deviate from the power law and hence cannot be collapsed (see deviations at small $s$, Fig.~\ref{Fig:MathSub}~D). These deviations decrease with increasing $\ps$, and with $\gamma \rightarrow 1^+$ \cite{Stumpf2005}, (\ref{sec:Exponent1}).  We call these deviations {\it ``hairs''} because they ``grow'' on the ``heads'' of the distribution as opposed to the tails of the distribution. In fact, the hairs allow to infer the system size from knowing the number of sampled units if the full systems exhibits a power-law distribution (\ref{sec:HairSize}).

In real world systems and in simulations, distributions often deviate from pure exponentials or pure power laws~\cite{Clauset2009,Levina2014}. We here treat the case that is typical for finite size critical systems, namely a power law that transits smoothly to an exponential around $s=s^\mathrm{cutoff}$ (e.g. Fig.~\ref{Fig:ModelSub}~A).
Under subsampling, $s_{\mathrm{sub}}^\mathrm{cutoff}$ depends linearly on the sampling probability: $s_{\mathrm{sub}}^\mathrm{cutoff}=\ps \cdot s^\mathrm{cutoff}$. Hence, the only solution to the power-law scaling relation (Eq.~\ref{Eq:ab}) that collapses (to the best possible degree), both, the power-law part of distributions \textit{and} the onsets of the cutoff is the one with $a=b=1$:
\begin{equation}
P(s)\approx\ps \Ps(\ps \cdot s). \label{Eq:SubsScal}
\end{equation}
As this scaling is linear in $p$, we call it p-scaling. A priori, p-scaling is different from the scaling for exponentials (Eq.~\ref{Eq:ExpScal}). 
However, p-scaling is a limit case of the scaling for exponentials under the assumption that $\lambda \ll \ps$: Taylor expansion around $\lambda=0$  
results in the scaling relation $P(s)\approx \ps \Ps(\ps \cdot s)$, i.e. the same as derived in Eq.~\ref{Eq:SubsScal}. 
Indeed,  for exponentials with $\lambda=0.001$ p-scaling  results in a nearly perfect collapse for all $\ps>0.01$, however $\ps\leq 0.01$ violates the $\lambda\ll p$ requirement and the collapse breaks down (Fig.~\ref{Fig:MathSub}~B, inset).  
Thus p-scaling collapses power laws with exponential tail if $\lambda$ is small, and also much smaller than the sampling probability. 
This condition is typically met in critical, but not in subcritical systems (Sec.~\ref{Suppl:Subcr}).


\subsection{Subsampling in critical models}

Experimental conditions typically differ from the idealized, mathematical formulation of subsampling derived above: Distributions do not follow perfect power laws or exponentials, and sampling is not necessarily binomial, but restricted to a fixed set of units. To mimic experimental conditions, we simulated avalanche generating models with fixed sampling in a critical state, because at criticality, subsampling effects are expected to be particularly strong: In critical systems, avalanches or clusters of activated units can span the entire system and thus under subsampling they cannot be fully assessed.


  {
We simulated critical models with different exponents of $P(s)$ to assess the generality of our analytically derived results. 
The first model is the widely used branching model (BM)~\cite{Priesemann2014,Watson1875, Harris1963, Haldeman2005, Larremore2012}, and the second model is the Bak-Tang-Wiesenfeld model (BTW)~\cite{Bak1987}, both studied in two variants. Both models display avalanches of activity after one random unit (neuron) has been activated externally (drive). In the BM, activity propagates stochastically, i.e.~an active neuron activates any of the other neurons with a probability $p_{\mathrm{act}}$. Here $p_{\mathrm{act}}$ is the control parameter, and the model is critical in the infinite size limit if one spike on average triggers one spike in its postsynaptic pool (see Methods). We simulated the BM  on a fully connected network and on a sparsely connected network. The avalanche size distributions of both BM variants have an exponent  $\approx 1.5$ \cite{Harris1963}, and for both variants, subsampling results are very similar (Sec.~\ref{sec:EHE_sparseBM}). Hence in the main text we show results for the fully connected BM, while the results for the sparsely connected BM are displayed, together with results of a third model, the non-conservative model from Eurich, Herrmann \& Ernst (EHE-model)~\cite{Eurich2002}, in Sec.~\ref{sec:EHE_sparseBM}. As expected, distributions of all critical models collapse under p-scaling.}

  {In the BTW, activity propagates deterministically via nearest neighbors connections. Propagation rules reflect a typical neural non-leaky ``integrate-and-fire'' mechanism: Every neuron sums (integrates) its past input until reaching a threshold, then becomes active itself and is reset. The BTW was implemented classically with nearest neighbor connections on a 2D grid of size $M=L \times L$ either with open (BTW), or with circular (BTWC) boundary conditions. For the BTW/BTWC the exponent of $P(s)$ depends on the system size, and for the size used here ($M=2^{14}$) it takes the known value of $\approx 1.1$~\cite{Lubeck_1997}. Thus the slope is flatter than 1.29, which is expected for the infinite size BTW ~\cite{Lubeck_1997, Munoz_1999}.}

For subsampling, $N$ units were pre-chosen randomly. This subsampling scheme is well approximated by binomial subsampling with $\ps=N/M$ in the BM, because the BM runs on a network with full or annealed connections, and hence units are homogeneously connected. In the BTW/BTWC, subsampling violates the binomial sampling assumption, because of the models' deterministic, local dynamics.

\begin{figure}
	\begin{center}
		{\includegraphics[width=0.7\textwidth]{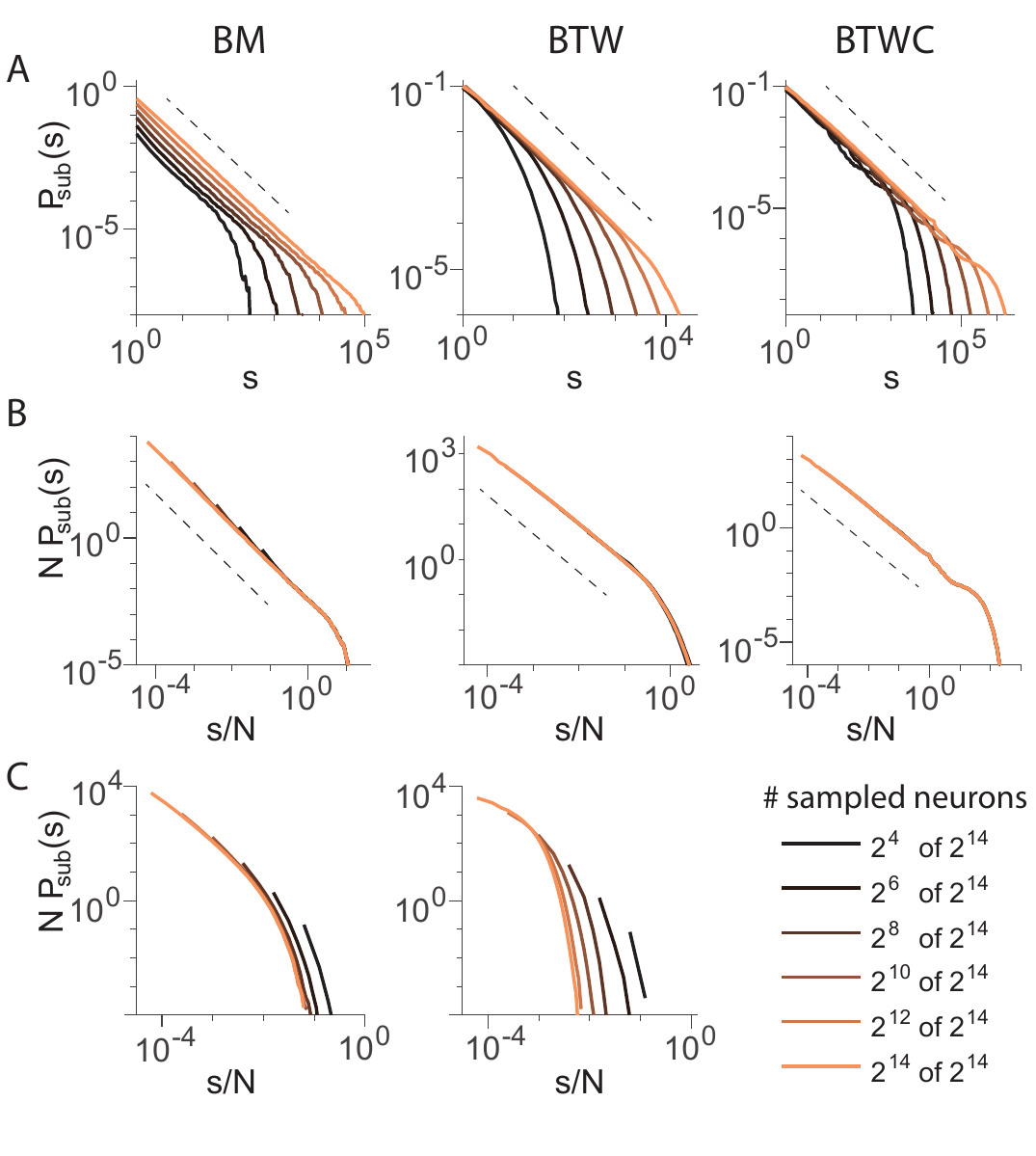}}
	\end{center}
	{\caption{Subsampling scaling in critical and subcritical models. The three columns show results for the branching  model (BM), the Bak-Tang-Wiesenfeld model (BTW), and the BTW with periodic boundary conditions (BTWC). 
	\textbf{A}: Avalanche size distribution $\Ps(s)$ for different degrees of subsampling, as denoted in the legend. 
	\textbf{B}: Same distributions as in A, but with p-scaling. (Note that scaling by $N$ leads to a collapse equivalent to scaling by $p=N/M$ at fixed system size $M$). 
	\textbf{C}: Scaled distributions from subcritical versions of the models. Here, results for the BTWC are extremely similar to those of the BTW and are thus omitted.
	Dashed lines indicate power-law slopes of $-1.5$ and $-1.1$ for the BM and BTW/BTWC, respectively, for visual guidance.}
	
	\label{Fig:ModelSub}}
	
\end{figure}


For all models, the avalanche distributions under full sampling  transit from an initial power law to an exponential at a cutoff $s^\mathrm{cutoff} \approx M$ due to finite size effects (Fig. \ref{Fig:ModelSub}~A). 
 For small $s$, the hairs appear in the BM, originating from subsampling power laws (Fig. \ref{Fig:ModelSub}~B, see Fig.~\ref{Fig:finite_size_vs_subsampling}~A for a flattened version). These hairs are almost absent in the BTW/BTWC, because the power-law slope is close to unity (\ref{sec:Exponent1}). The tails, even those of the BTWC, which have an unusual transition at the cutoff, collapse well. 
The BTWC is an exception in that it has unusual finite size effects, translating to the characteristic tails of $P(s)$. In fact, here the tails collapse better when applying fixed instead of binomial subsampling (Fixed subsampling refers to pre-choosing a fixed set of units to sample from; this may violate mean-field assumptions). This is because loosely speaking, binomial subsampling acts as a low pass filter on $P(s)$, smearing out the peaks, while fixed subsampling conserves the shape of the tails better here, owing to the compactness of the avalanches specifically in the 2D, locally connected BTWC. 
Overall, despite the models' violation of mean-field assumptions, the analytically motivated p-scaling ansatz allows to infer $P(s)$ from subsampling, including the detailed shapes of the tail.

\subsection{Distinguishing critical from subcritical systems}

Distinguishing between critical and subcritical systems under subsampling is particularly important when testing the popular hypothesis that the brain shows signatures of ``critical dynamics''. Criticality is a dynamical state that  maximizes information processing capacity in models, and therefore is a favorable candidate for brain functioning~\cite{Haldeman2005,Bertschinger2004,Boedecker2012,Shew2013}. Typically, testing for criticality in experiments is done by assessing whether the ``neural avalanche'' distributions follow power laws~\cite{Beggs2003}. Here, subsampling plays a major role, because at criticality avalanches can propagate over the entire network of thousands or millions of neurons, while millisecond precise sampling is currently constrained to about 100 neurons. Numerical studies of subsampling reported contradictory results~\cite{Priesemann2009, Priesemann2014, Ribeiro2010, Yu2014, Ribeiro2014}. Therefore, we revisit subsampling with our analytically derived scaling, and compare scaling for critical and subcritical states.

In contrast to critical systems, \textit{subcritical} ones lack large avalanches, and the  cutoff of the avalanche size distribution is independent of the system size (if $M$ is sufficiently large; Fig.~\ref{Fig:SubExpTail}). As a consequence, the distributions obtained under subsampling do not collapse under p-scaling (Fig.~\ref{Fig:ModelSub}~C). In fact, there exists no scaling that can collapse all subsampled distributions (for any $\ps$) simultaneously, as outlined below, and thereby p-scaling can be used to distinguish  critical from non-critical systems. 

The violation of p-scaling in subcritical systems arises from the incompatible requirement for scaling at the same time the power-law part, the exponential tail, and the cutoff onset $s_\mathrm{sub}^\mathrm{cutoff}$ . On the one hand, the exponential tail becomes increasingly steeper with distance from criticality (larger $\lambda$), so that the relation $\lambda \ll \ps$ required for p-scaling (Eq.~\ref{Eq:SubsScal}) does not hold anymore for small $\ps$ (Sec.~\ref{Suppl:Subcr}). 
Thus, a collapse of the tails would require the scaling ansatz for exponentials (Eq.~\ref{Eq:ExpScal}).
On the other hand, slightly subcritical models still exhibit power-law behavior up to a
 cutoff $s^\mathrm{cutoff}:=c$ that is typically much smaller than the system size ($c \ll M$). To properly scale this part of the distribution, p-scaling is required. Likewise, the onset of the cutoff scales under subsampling with $\ps$: $s_\mathrm{sub}^\mathrm{cutoff} = c \cdot \ps$, requiring a scaling of the $s$-axis in the same manner as in the p-scaling. Thus, because the exponential decay requires different scaling than the power law and $s_\mathrm{sub}^\mathrm{cutoff}$, no scaling ansatz can collapse the entire distributions from ``head to tail''.


\subsection{  {Cluster definition, binning, and subsampling scaling} }
\label{sec:binning}

	\begin{figure}
		\centering
		\includegraphics[width=.89\linewidth]{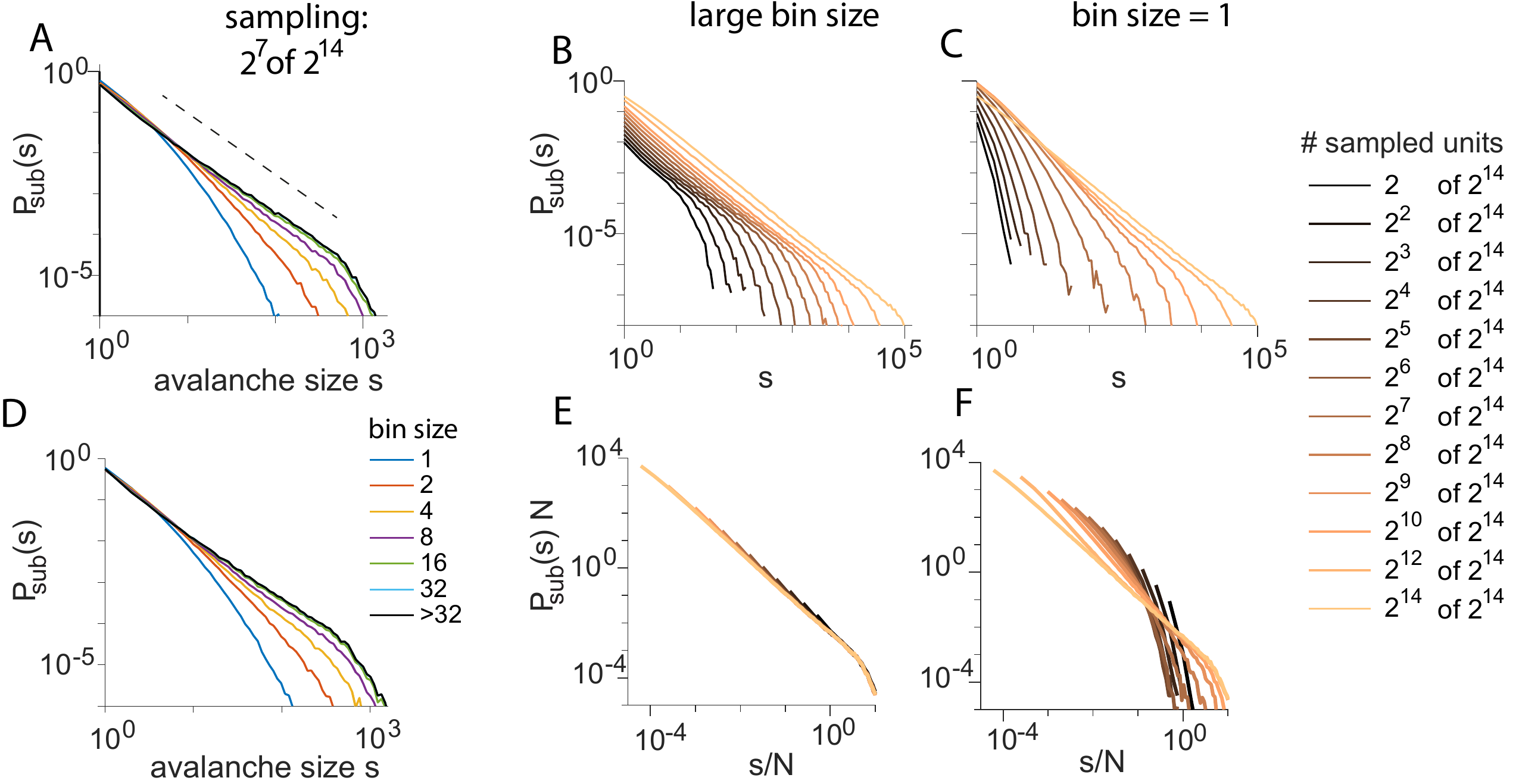}
		\caption{{Impact of binning on avalanche size distributions $\Ps(s)$ and scaling. 
				\textbf{A, D}: Sampling $N=2^7$ units at different bin sizes from sparse (A) and fully-connected network (D). For small bin sizes ($< 16$ steps), $\Ps(s)$ deviates from a power law with slope $1.5$ (dashed line). For larger bin sizes ($\geq 32$ steps), $\Ps(s)$ is bin size invariant and shows the expected power law with cutoff.
				\textbf{B, E}: Same as Fig.~\ref{Fig:ModelSub}; for sufficiently large bin sizes $\Ps(s)$ collapsed under subsampling scaling. 
				\textbf{C, F}: When applying a small bin size, here 1 step,
				  $\Ps(s)$ does not collapse. 
				Parameters: Critical branching model (BM) with size $M=2^{14}$ and sparse connectivity ($k=4$),
  {except for D that has all-to-all connectivity ($k=M$)}. 
			}}\label{Fig:AvBinning}
		\end{figure}

The main focus of this paper is to show how distributions of avalanches, node degrees or other ``clusters'' change under spatial subsampling, and how to infer the distribution of the fully sampled system from the subsampled one. 
To this end, it is essential that the clusters are extracted unambiguously, i.e. one cluster in the full system translates to exactly one cluster (potentially of size zero) under subsampling. This condition is easily realized for the degree of a node: One simply takes into account only those connections that are realized with other \textit{observed} nodes. For avalanches, this condition can also be  fulfilled easily if the system shows a separation of time scales (STS), i.e. the pauses between subsequent avalanches are much longer than the avalanches themselves (see Sec.~\ref{sec:avdef}).  
With a STS, temporal binning~\cite{Beggs2003} can be used to unambiguously extract avalanches under subsampling. However, the chosen bin size must neither be too small nor too large: If too small, a single avalanche on the full system can be ``cut'' into multiple ones when entering, leaving, and re-entering the recording set. 
  {This leads to steeper $\Ps(s)$ with smaller bin size (Fig.~\ref{Fig:AvBinning}\,A).}
In contrast, if the bin size is too large, subsequent avalanches can be ``merged'' together. For a range of intermediate bin sizes, however, $\Ps(s)$ is invariant to changes in the bin size. 
{In Fig.~\ref{Fig:AvBinning}\,A, the invariance holds for all bin sizes $32< \mathrm{bin\: size} <\infty$}.   { The result does not depend on the topology of the network (compare Fig.~\ref{Fig:AvBinning}\,A for a network with sparse topology and Fig.~\ref{Fig:AvBinning}\,D for fully connected network).}
  {If a system, however, lacks a STS, then $\Ps(s)$ is expected to change for any bin size. This may underlie the frequently observed changes in $\Ps(s)$ in neural recordings~\cite{Priesemann2009,  Priesemann2013, Beggs2003, Arviv2015,  Shriki2013, Hahn2010}, as discussed in~\cite{Priesemann2014}.} 

To demonstrated the impact of the bin size on p-scaling, we
here used the branching model (BM), which has a full STS, i.e. the time between subsequent avalanches is mathematically infinite. 
When sampling $N=2^7$ out of the $M=2^{14}$ units, then $\Ps(s)$ deviates from a power law for small bin sizes and only approaches a power law with the expected slope of $1.5$ for bin sizes larger than 8 steps (Fig.~\ref{Fig:AvBinning}\,A). The same holds for subsampling of any $N$: With sufficiently large bin sizes, $\Ps(s)$ shows the expected approximate power law (Fig.~\ref{Fig:AvBinning}\,B). In contrast, for small bin sizes avalanches can be cut, and hence $\Ps(s)$ deviates from a power law (Fig.~\ref{Fig:AvBinning}\,C).   {This effect was also observed in~\cite{Priesemann2014,Ribeiro2014}, where the authors used small bin sizes and hence could not recover power laws in the critical BM under subsampling, despite a STS.}
Thus in summary, p-scaling only collapses those $\Ps(s)$, where avalanches were extracted unambiguously, i.e. a sufficiently large bin size was used (compare Fig.~\ref{Fig:AvBinning}\,E and F).

The range of bin sizes for which $\Ps(s)$ is invariant depends on the specific system. For the experiments we analyzed in the following section, we found such an invariance for bin sizes from 0.25 ms to 8 ms if $\Ps(s)$ follows a power law, indicating indeed the presence of a STS (Fig.~\ref{fig:PLscal}\,D). Thus our choice of 1 ms bin size suggests an unambiguous extraction of avalanches, and in this range p-scaling works as predicted theoretically.


\subsection{Subsampled neural recordings: Learning more by sampling less}

We applied p-scaling to neural recordings of developing networks \textit{in vitro} to investigate whether their avalanches indicated a critical state. To this end, we evaluated recordings from $N=58$ multi-units (see Methods, \cite{Wagenaar2006}). This is only a small fraction of the entire neural network, which comprised $M\approx 50.000$ neurons, thus the avalanche size distribution obtained from the whole analyzed data is already a subsampled distribution $\Ps(s)$. To apply p-scaling, we generated a family of distributions by further subsampling, i.e. evaluating a subset $N'<N$ of the recorded units. In critical systems, p-scaling is expected to collapse this family of distributions if avalanches are defined unambiguously, as outlined above (Sec.~\ref{sec:binning}).

\begin{figure}
	\centering
	\includegraphics[width=1.0\linewidth]{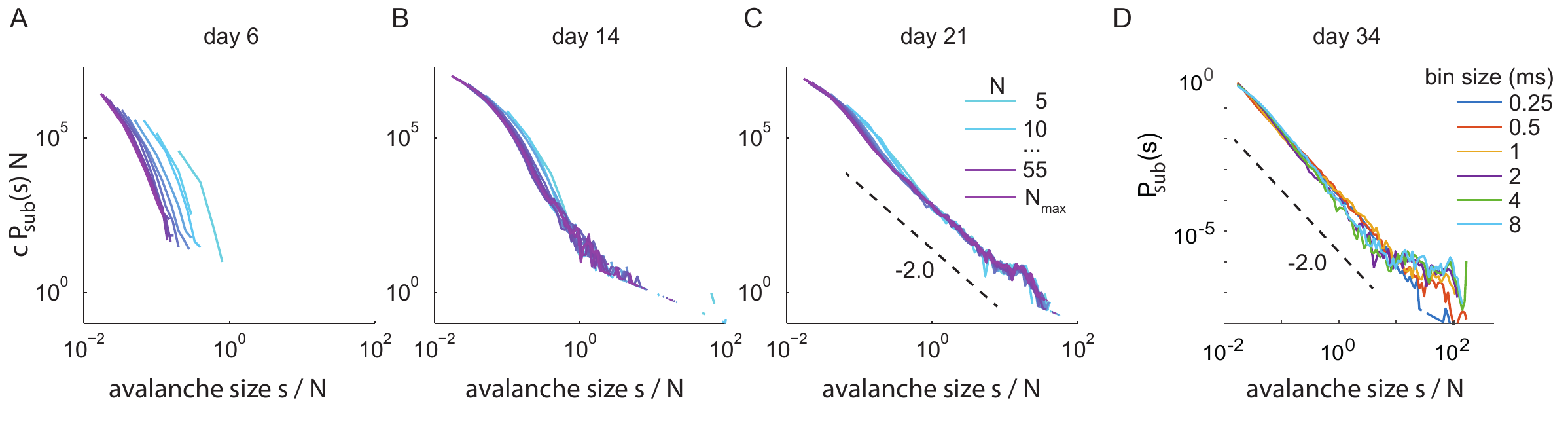}
	\caption{Avalanche size distributions $c\cdot\Ps(s)$ (in absolute counts) of spiking activity of developing neural networks \textit{in vitro}. 
	\textbf{A}: For young cultures, $\Ps(s)$ did not collapse under p-scaling, indicating that the full network does not show a power-law distribution for $\Ps(s)$. 
	\textbf{B,C}: More mature networks show a good collapse, allowing to extrapolate the 
distribution of the full network (see Fig.~\ref{Fig:SpikeFSall} for all recording days of all experiments). 
  {In panels A, B, and C the bin size is 1 ms, and $c$ is a total number of recorded avalanches in the full system, in A: $c=53,803$, in B: $c=307,908$, in C: $c=251,156$.} The estimated number of neurons in the cultures is $M\approx50,000$. 
{\textbf{D}: $\Ps(s)$ from sampling spikes from all electrodes but evaluated with different bin sizes (see legend); the approximate invariance of $\Ps(s)$ against changes in the bin size indicates a separation of time scales in the experimental preparation.}
} \label{fig:PLscal}
\end{figure}

Interestingly, for early stages of neural development, p-scaling does not collapse $\Ps(s)$, but for the more mature networks we found a clear collapse (Fig. \ref{fig:PLscal}; for all experiments see Fig.~\ref{Fig:SpikeFSall}). 
Thus developing neural networks start off with collective dynamics that is not in a critical state, but with maturation approach criticality~\cite{Tetzlaff2010, Pasquale2008}. 
Some of the mature networks show small bumps in $\Ps(s)$ at very large avalanche sizes ($s\approx 5000 \Leftrightarrow s/N \approx 60$). These very large avalanches comprise only a tiny fraction of all avalanches (about 2 in 10,000). At first glance, the bumps are reminiscent of supercritical systems. However, supercritical neural models typically show bumps at system or sampling size ($s=N$), not at those very large sizes. We discuss this in more detail in Sec.~\ref{Sec:SI_Experiments}, and suggest that the bumps are more likely to originate from neurophysiological finite size effects.

For the full, mature network, our results predict that $P(s)$ would extend not only over three orders of magnitude as here, but over six, because $p\approx 10^{-3}$ (Sec.~\ref{Sec:SI_Experiments}).
Our analysis of neural recordings illustrates how further spatial subsampling allows to infer properties of the full system, even if only a tiny fraction of its collective dynamics has been observed, simply by sampling even less ($N'<N$) of the full system.

\subsection{Subsampling versus finite size scaling} \label{S:sub_vs_fss}

\begin{figure}
	\centering
	\includegraphics[width=0.8\linewidth]{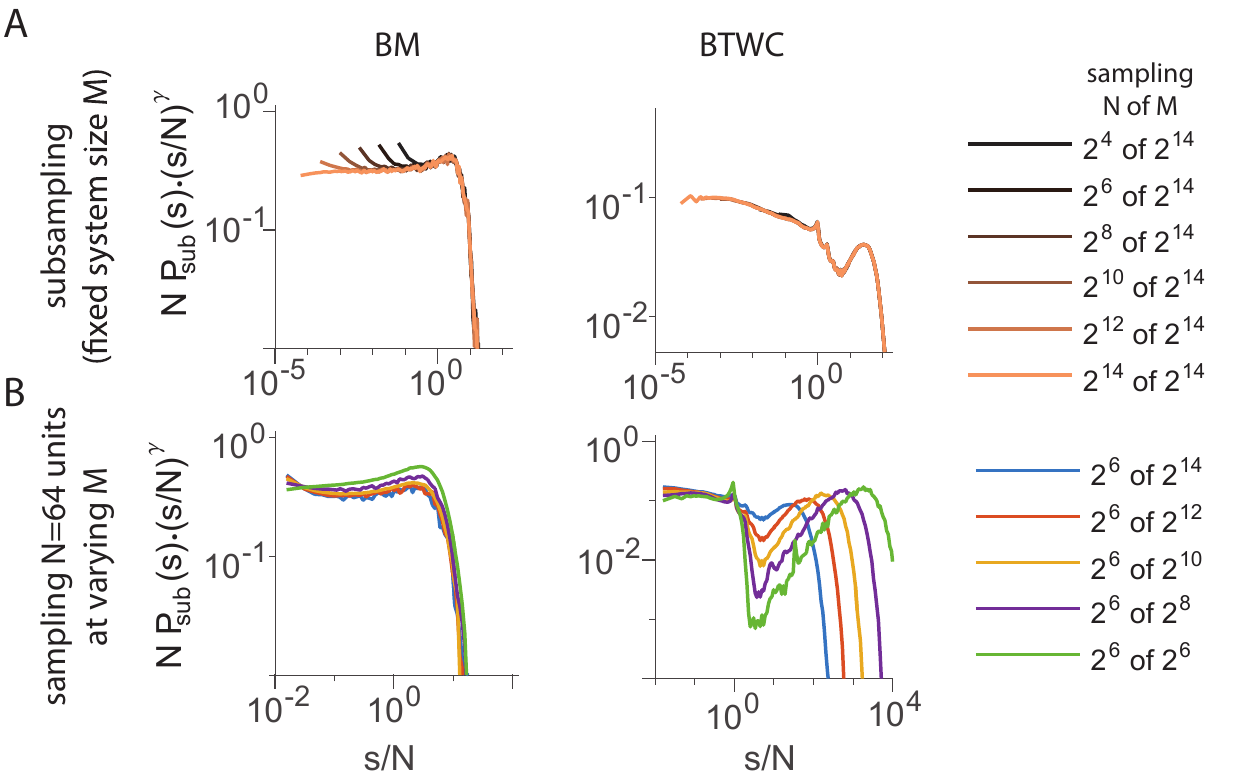}
	\caption{Changes in $\Ps(s)$ mediated by system size ($M$) and sampling size ($N$).  
	\textbf{A, B}: Scaled and flattened avalanche size distribution ($\Ps(s)$) for the branching model (BM, left) and the Bak-Tang-Wiesenfeld model with circular boundary conditions (BTWC, right); flattening is achieved by multiplying $\Ps(s)$ with a power law with appropriate slope $\gamma$.
	  {We used $\gamma=1.5$ and $\gamma=1$ for the BM and BTWC, respectively}.
	\textbf{A}: $\Ps(s)$ for different samplings ($N=2^4 \dots 2^{14}$) from models with fixed  size $M=2^{14}$. Note the ``hairs'' in the BM induced by subsampling. 
	\textbf{B}: $\Ps(s)$ from sampling a fixed number of $N=2^6$ neurons from models of different sizes ($M=2^6\dots 2^{14}$). Note the difference in distributions despite the same number of sampled neurons, demonstrating that finite size effects and subsampling effects are not the same.
}\label{Fig:finite_size_vs_subsampling}
\end{figure}

In the real world we are often confronted with data affected by both subsampling and finite system size effects, i.e. observations originated from a small part of a large, but not infinite system. 
Thus we need to deal with a combination of both: subsampling effects as a result of incomplete data acquisition and finite-size effects inherited from the full system. To disentangle influences from system size and system dynamics, finite size scaling (FSS) has been introduced~\cite{Privman1990,Levina2014a}. It allows to infer the behavior of an infinite system from a set of finite systems. 
At a first glance, finite size and subsampling effects may appear to be very similar. However, if they were, then distributions obtained from sampling $N$ units from any system with $N\leq M$ would be identical, i.e. independent of $M$. This is not the case, as e.g. the distributions for fixed $N=2^6$ clearly depend on $M$ (Fig.~\ref{Fig:finite_size_vs_subsampling}~B).
In fact, in both models the tails clearly inherit signatures of the full system size. Moreover, in the BM, subsampling a smaller fraction $\ps=N/M$ of a system increases the ``hairs'', an effect specific to subsampling, not to finite size (see the increasing convexity of the flat section with decreasing $\ps$ in the BM, Fig.~\ref{Fig:finite_size_vs_subsampling}~B).

\begin{figure}
		\centering
		\includegraphics[width=.95\linewidth]{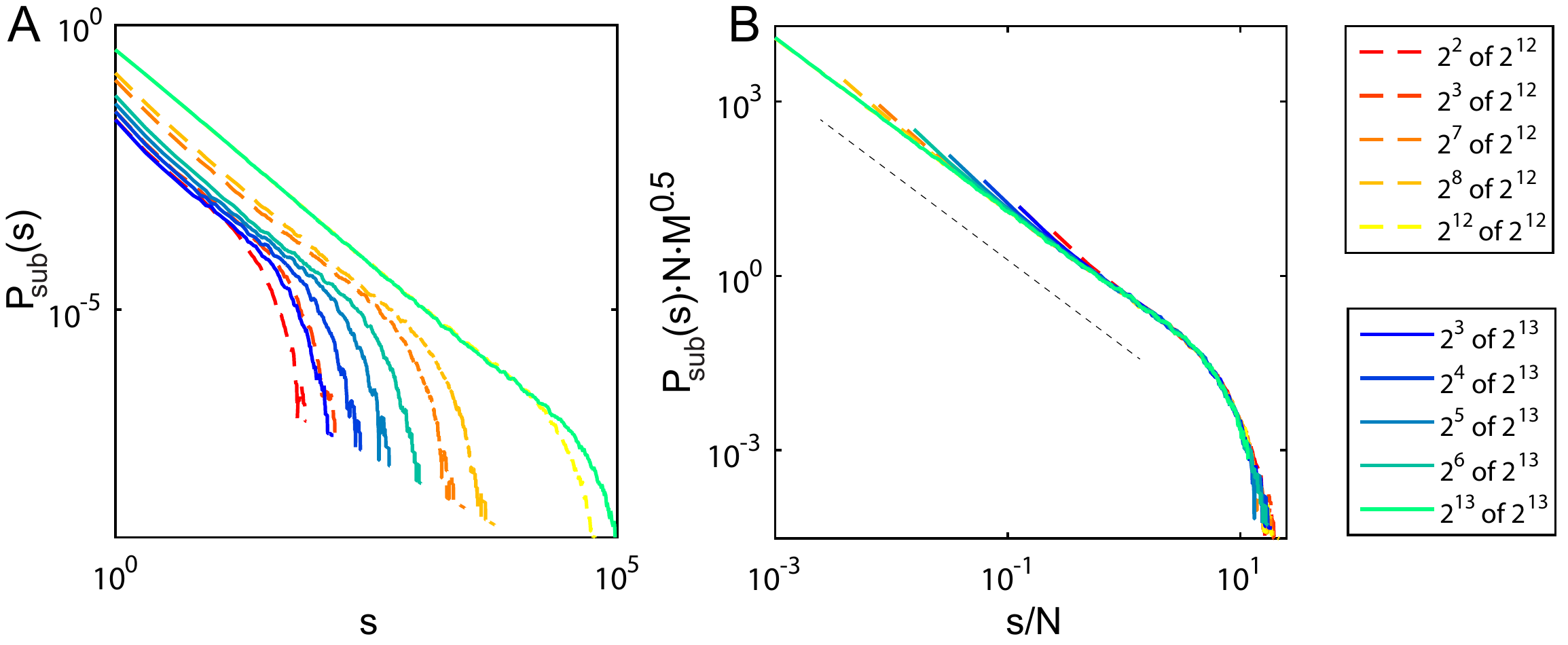}
		\caption{   Subsampling scaling combined with finite-size scaling. 
		\textbf{A}: Subsampling distributions for different numbers of sampled units $N$ from the BM with system sizes $M_1=2^{12}$ and $M_2=2^{13}$. 
		\textbf{B}: Distributions collapsed as predicted by applying subsampling-finite-size scaling (Eq.~\ref{Eq:fss_ss}, with $\gamma=1.5$ and $\nu=1$).
		The dashed black line indicates a slope of $-1.5$ for visual guidance.}
		\label{Fig:FSS_subs}
	\end{figure}

Importantly, as shown above, for critical systems one can always scale out the impact of subsampling, and thereby infer the distribution of the full system, including its size specific cutoff shape (Fig.~\ref{Fig:finite_size_vs_subsampling}~A). 
{   Hence, it is possible to combine FSS and subsampling scaling  (detailed derivation are in Sec.~\ref{Sec:FSS_sub}): Consider a critical system, where FSS is given by: $M^\beta P(s M^\nu;M) = g(s),$ here $g(s)$ is a universal scaling function. Then FSS can be combined with subsampling scaling to obtain a universal subsampling-finite-size scaling:
	\begin{equation}
	N M^{\beta-1} \Ps(s N M^{\nu-1};M,N)=g(s). \label{Eq:fss_ss}
	\end{equation}
 Using Eq.~\ref{Eq:fss_ss} allows to infer the distribution for arbitrary subsampling ($N$) of any system size ($M$), Fig.~\ref{Fig:FSS_subs}. 

 }

\section{Discussion}

The present study analytically treats subsampling scaling for power laws (with cutoff), exponential distributions, and negative and positive binomial distributions.
For all other distributions, utmost care has to be taken when aiming at inferences about the full system from its subsampling. One potential approach is to identify a scaling ansatz numerically, {   i.e. 
minimizing the distance between the different $\Ps(s)$ numerically, in analogy to the approach for avalanche shape collapse~\cite{Priesemann2013, Shaukat2016,  Arviv2015, Marshall2016, Friedman2012}. We found that for our network simulations such a numerical approach identified the same scaling parameters as our analytic derivations~(Sec.~\ref{Sec:NumApproach}).}  However, given the typical noisiness of experimental observations, a purely numerical approach should be taken with a grain of salt, 
as long as it is not backed up by a circular form analytical solution.


Our analytical derivations assumed annealed sampling, which in simulations was well approximated by pre-choosing a random subset of neurons or nodes for sampling.  Any sampling from randomly connected networks is expected to lead to the same approximation. However, in networks with e.g.~local connectivity, numerical results depend strongly on the choice of sampled units~\cite{Priesemann2009}.
For example, for windowed subsampling (i.e. sampling a local set of units) a number of studies reported strong deviations from the expected power laws in critical systems or scale free networks~\cite{Priesemann2009, Ribeiro2010, Gerhard2011}. 
In contrast, random subsampling, as assumed here for our analytical derivations, only leads to minor deviations from power laws (hairs). Thus to diminish corruption of results by subsampling, future experimental studies on criticality should aim at implementing random instead of the traditional windowed sampling, e.g. by designing novel electrode arrays with pseudo-random placement of electrodes on the entire area of the network. In this case, we predict deviations from power laws to be minor, i.e. limited to the ``hairs'' and the cutoff.

We present here first steps towards a full understanding of subsampling. With our analytical, mean-field-like approach to subsampling we treat two classes of distributions and explore  corresponding simulations. In future, extending the presented approach to a window-like sampling, more general forms of correlated sampling, and to further classes of distributions will certainly be of additional importance to achieve unbiased inferences from experiments and real-world observations.


\section{Methods}

\subsection{Analytical derivations}
The analytical derivations are detailed in the Supplementary Information.

\subsection{Simulations}
We simulated two models, the Bak-Tang-Wiesenfeld Model (BTW) with open and with circular (i.e. periodic) (BTWC) boundary conditions, and the branching model (BM) with full and with annealed sparse connectivity. 

\subsubsection{Bak-Tang-Wiesenfeld Model}
The Bak-Tang-Wiesenfeld Model (BTW) \cite{Bak1987}, was realized on a 2D grid of $L\times L=M$ units, each unit connected to its four nearest neighbors. Units at the boundaries or edges of the grid  have either 3 or 2 neighbors, respectively (open boundary condition). Alternatively,  the boundaries are closed circularly, resulting in a torus (circular or periodic boundary condition, BTWC). 
Regarding the activity, a unit at the position $(x,y)$ carries a potential $z(x,y,t)$  at time $t$, ($z,t \in \mathds{N}_0$). If $z$ crosses the threshold of 4 at time $t$, its potential is redistributed or ``topples'' to its nearest neighbors:

\begin{align*}
	\text{if } z(x,y,t) \geq 4: & \\
	& z(x,y,t+1)  = z(x,y,t) - 4 \\
  &	z(x\pm 1,y\pm 1,t+1)  = z(x\pm 1,y\pm 1,t) + 1
\end{align*}

$z(x\pm1,y\pm1)$ refers to the 4 nearest neighbors of $z(x,y)$. The BTW/BTWC is in an absorbing (quiescent) state if $z(x,y)<4$, for all $(x,y)$. From this state, an ``avalanche'' is initiated by setting a random unit $z(x,y)$ above threshold: $z(x,y,t+1) = z(x,y,t)+4$. The activated unit topples as described above and thereby can make neighboring units cross threshold. These in turn topple, and this toppling cascade propagates as an avalanche over the grid until the model reaches an absorbing state. The size $s$ of an avalanche is the total number of topplings. Note that the BTW/BTWC are initialized arbitrarily, but then run for sufficient time to reach a stationary state. Especially in models with large $M$ this can take millions of time steps.

The BTW and the BTWC differ in the way how dissipation removes potential from the system. Whereas in the BTW potential dissipates via the open boundaries, in the BTWC an active unit is reset without activating its neighbors with a tiny probability, $\pd=10^{-5}$. For BTW an additional dissipation in a form of small $\pd$ can be added to make the model subcritical.

\subsubsection{Branching model}
The branching model (BM) corresponds to realizing a classical branching process on a network of units~\cite{Priesemann2014, Harris1963, Haldeman2005}. In the BM, an avalanche is initiated by activating one unit. This unit activates each of the $k$ units it is connected to with probability $p_{\mathrm{act}}$ at the next time step. These activated units, in turn, can activate units following the same principle. This cascade of activations forms an avalanche which ends when by chance no unit is activated by the previously active set of units. The control parameter of the BM is $\sigma = p_{\mathrm{act}}\cdot k$. For $\sigma=1$, the model is critical in the infinite size limit.
 We implemented the model with full connectivity ($k=M$) and with sparse, annealed connectivity ($k=4$).
The BM can be mathematically rigorously associated with activity propagation in an integrate and fire network~\cite{LevinaDiss,Levina2007d}.

  {For implementation of the BM with full connectivity ($k=M=2^{14}$), note that the default pseudo-random number generator (PRNG) of Matlab(R) (R2015b) can generate avalanche distributions that show strong noise-like deviations from the expected power-law distribution. These deviations cannot be overcome by increasing the number of avalanches, but by specifying a different PRNG. We used the ``Multiplicative Lagged Fibonacci'' PRNG for the results here, because it is fairly fast.}

\subsubsection{Subcritical models}
To make the models subcritical, in the BM $\sigma$ was set to $\sigma=0.9$, and in the BTW/BTWC the dissipation probability $\pd$ was set to  $\pd = 0.1$, which effectively corresponds to $\sigma=0.9$, because 90\% of the events are transmitted, while 10\% are dissipated.

{\subsubsection{Avalanche extraction in the models}
\label{sec:avdef}
The size $s$ of an avalanche is defined as the total number of spikes from the seed spike until no more units are active. Under subsampling, this translates to the total number of spikes that occur on the pre-chosen \textit{set of sampled units} (fixed subsampling). In principle, the avalanches could also have been extracted using the common binning approach~\cite{Beggs2003}, as all the models  were simulated with a separation of time scales (STS), i.e. the time between subsequent avalanches is by definition much longer than the longest-lasting avalanche. Hence applying any bin size that is longer than the longest avalanche, but shorter than the pauses between avalanches would yield the same results for any subsampling.
}

\subsection{Neural recordings}

\subsubsection{Data acquisition and analysis}

The spike recordings were obtained by Wagenaar et al.~\cite{Wagenaar2006} from an \textit{in vitro} culture of $M\approx 50,000$ cortical neurons. Details on the  preparation, maintenance and recording setting can be found in the original publication. In brief, cultures were prepared from embryonic E18 rat cortical tissue. Recording duration of each data set was at least 30 min. The recording system comprised an $8\times8$ array of 59 titanium nitride electrodes with 30 $\mu$m diameter and 200 $\mu$m inter-electrode spacing, manufactured by Multichannel Systems (Reutlingen, Germany). As described in the original publication, spikes were detected online using a threshold based detector as upward or downward excursions beyond 4.5 times the estimated RMS noise~\cite{Wagenaar2005}. Spike waveforms were stored, and used to remove duplicate detections of multiphasic spikes. Spike sorting was not employed, and thus spike data represent multi-unit activity.

For the spiking data, avalanches were extracted using the classical binning approach as detailed in~\cite{Beggs2003, Priesemann2014}. In brief, temporal binning is applied to the combined spiking activity of all channels. Empty bins by definition separate one avalanche from the next one. The avalanche size $s$ is defined as the total number of spikes in an avalanche. The bin size applied here was 1 ms, because this reflects the typical minimal time delay between a spike of a driving neuron and that evoked in a monosynaptically connected receiving neuron, and because 1 ms is in the middle of the range of bin sizes that did not change the avalanche distribution $\Ps(s)$ (Sec.~\ref{sec:binning}, Fig.~\ref{fig:PLscal}~D).

Application of p-scaling by definition requires that one avalanche in the full system translates to one avalanche (potentially of size zero) under subsampling, i.e. an avalanche must not be ``cut'' into more than one, e.g. when leaving and re-entering the recording set. 
	This can be achieved in experiments that have a separation of time scales by applying a sufficiently large bin size, because this allows for an unambiguous avalanche extraction~\cite{Priesemann2014} (Sec.~\ref{sec:binning}). Indeed, the \textit{in vitro} recordings we analyze here appear to show a separation of time scales: We found that varying the applied bin size around 1 ms hardly changed $\Ps(s)$ (Fig.~\ref{fig:PLscal}~D). In contrast, using too small bin sizes would have led to ``cutting'' avalanches, which impedes the observation of power laws, and consequently prevents the collapse (illustrated for the BM, Fig.~\ref{Fig:AvBinning}).

\subsubsection{Data availability and selection criteria}

We evaluated 10 recordings for each day, because then the na\"ive probability of finding the expected behavior consistently in all of them by chance is at most $p=(1/2)^{10} <0.001$. The experimental data was made available online by the Potter group~\cite{Wagenaar2006}. In detail, we downloaded from the “dense” condition the \textit{in vitro} preparations 2-1, 2-3, 2-4, 2-5, 2-6, 6-1, 6-2, 6-3, 8-1, 8-2, 8-3, and for each preparation one recording per week (typically days 7, 14, 21, 28, 34/35, but for some experiments one or two days earlier), except for experiment 6-2 where we only got the first three weeks, and 6-3 where we got the last two weeks. We analyzed and included into the manuscript all recordings that we downloaded.

{  
\subsection*{Acknowledgments and Funding}
The authors thank Georg Martius and Michael Wibral for inspiring discussions and support. We thank Jens Wilting, Johannes Zierenberg, and Jo\~ao Pinheiro Neto for careful proof reading.
AL received funding from the People Program (Marie Curie Actions) of the 
European Union's Seventh Framework Program (FP7/2007-2013) under REA grant 
agreement no.~[291734].
VP received financial support from the Max Planck Society.
AL and VP received financial support from the German Ministry for Education and Research (BMBF) via the Bernstein Center for Computational Neuroscience (BCCN) G\"ottingen under Grant No. 01GQ1005B.

\subsection*{Author contributions}
AL and VP contributed equally to the work. All results were developed jointly. 
}

%% file: SI.tex
\pagebreak
\section{Supplementary Information for ``Subsampling scaling: a theory about inference from partly observed systems''}

In the following, we derive in detail the novel subsampling scaling. We first introduce the definition and basic results of subsampling in analogy to Stumpf et al.~\cite{Stumpf2005}, who treated subsampling of graphs.  {   We then extend the aforementioned study as follows:
	\begin{itemize}
		\item First, we focus on an analytical inference of the distribution of the full system from the subsampling, a topic that was not touched by Stumpf et al. To this end we derive (a) the exact subsampling scaling for negative binomials and exponentials, and (b) the approximate scaling for power-law distributions. 
		\item Second, we explicitly show how to derive the system size from subsampling induced deviations from power laws (``hairs'').
		\item Third, we treat the relation between subsampling and finite size effects.
		\item Last, we apply our analytically derived subsampling scaling ansatz to infer the probability distribution of avalanche sizes in developing neural networks.
			\end{itemize}  
		}

\subsection{Mathematical subsampling}

Let $X$ be a discrete, non-negative random variable with probability distribution $P(X=s)=P(s)$, with $s \in \mathbb{N}_0$, then  $G(z)=\sum_{s=0}^{\infty} z^s P(s)$ is the 
corresponding probability generating function (PGF).   For distributions such as power laws, where $s=0$ is not supported,  $s$ is constrained to $s \in \mathbb{N}$, and the probability distribution needs to be normalized accordingly. 
$X$ represents the size of a set of ``events'' that comprise a ``cluster'', e.g. the number of spikes in an avalanche, or the degree of a node. 
For subsampling, we assume that each of the events in the cluster is sampled independently with probability $\ps$, resulting in a random variable for the observed cluster size, $\Xs$ \cite{Stumpf2005}. Thus the probability $\Ps(\Xs=s)$ to observe a cluster of size $s$ is derived using a binomial distribution:

\begin{equation*}
\Ps(s)=\sum_{k = s}^\infty P(k) {k \choose s}\ps^s (1-\ps)^{k-s}.
\end{equation*}

{  The PGF $G_\mathrm{sub}(z;\ps)$ for $\Xs$ with given $\ps$ is thus:}
\begin{eqnarray*}
\Gs(z;\ps) & = &\sum_{s=0}^{\infty} z^s \Ps(s) \\
& = &\sum_{s=0}^{\infty} z^s \sum_{k = s}^\infty P(k) {k \choose s}\ps^s (1-\ps)^{k-s}\\
& = &\sum_{k=0}^{\infty}  P(k) \sum_{s=0}^k z^s   {k \choose s}\ps^s (1-\ps)^{k-s} \\
& = &\sum_{k=0}^{\infty} P(k) (z \ps + (1-\ps))^k,
\end{eqnarray*}

Thus the PGF of $X$ and $\Xs$ show a direct relation~\cite{Stumpf2005}:
\begin{equation}
\Gs(z;\ps)=G(1-\ps (1-z)) \label{Eq:PGFsub}
\end{equation}
As a consequence, the expected values of $X$ and $\Xs$ also are closely related:
Using the expression for the expected value of $X$, $E(X)=G'(1^-)$
\begin{equation}
E(\Xs)=\Gs'(1^-;\ps)=G'(1-\ps (1-1^-))=\ps G'(1^-)=\ps E(X)
\end{equation}
 
These relations hold for any $P(s)$, however, only for specific $P(s)$, namely positive and negative binomials, the full and subsampled system's $P(s)$ follow the same family of distributions~\cite{Stumpf2005}, e.g. if $P(s)$ is a binomial distribution, then $P_\mathrm{sub}(s)$ also is a binomial, but with different parameters.

\subsubsection{Subsampling of negative binomial and exponential distributions} \label{sec:SupplExp}
Assuming that $X$ follows a negative binomial distribution $X\sim  {\mathrm{NB}}(r,p_{ {\mathrm{NB}}})$,

\begin{equation}
P(X=s)={{s+r-1 \choose s} p_{ {\mathrm{NB}}}^r (1-p_{ {\mathrm{NB}}})^s},
\end{equation}

then the expectation of $X$ is given by

\begin{equation*}
m = E(X)=r\frac{(1-p_{ {\mathrm{NB}}})}{p_{ {\mathrm{NB}}}}, 
\end{equation*}

and the PGF is

\begin{equation*}
G(z)
= \left(\frac{p_{ {\mathrm{NB}}}z-z+1}{p_{ {\mathrm{NB}}}} \right)^{-r}
= \left(1+\frac{m}{r}(1-z)\right)^{-r}.
\end{equation*}

Using Eq.~\ref{Eq:PGFsub} then returns the PGF under subsampling,

\begin{equation*}
\Gs(z;\ps)=\left(1+\frac{\ps m}{r}(1-z)\right)^{-r},
\end{equation*}

which corresponds to the negative binomial distribution with the same $r$, but different $p_{ {\mathrm{NB}}}$, selected such that

\begin{equation}
\frac{1-p_{ {\mathrm{NB}}}'}{p_{ {\mathrm{NB}}}'}=\ps \frac{1-p_{ {\mathrm{NB}}}}{p_{ {\mathrm{NB}}}}. 
\label{Eq:NegBinSub}
\end{equation}

In the special case of $r=1$, the negative binomial is a geometric distribution with probability parameter $p_{ {\mathrm{NB}}}$, and the discrete exponential distribution is a particular parametrization of the geometric distribution $1-p_{ {\mathrm{NB}}}=e^{-\lambda}$.

\begin{equation*}
P_{\mathrm{exp}}(s)= (1-e^{-\lambda})\cdot e^{-\lambda s}, \: \mathrm{with} \: s \in \mathds{N_0}.
\end{equation*}

Using equation~\ref{Eq:NegBinSub}, the relation between $\lambda$ and $\lams$ is:

\begin{equation*}
\frac{e^{-\lams}}{1-e^{-\lams}}=\ps \frac{e^{-\lambda}}{1-e^{-\lambda}} \: \Leftrightarrow  \:
e^{\lams}-1=\frac{e^{\lambda}-1}{p}.
\end{equation*}

Solving this equation with respect to $\lams$ we obtain:

\begin{equation}
\lams=\ln\left( \frac{e^\lambda+\ps-1}{\ps}\right).
\end{equation}

\subsubsection{Subsampling of power-law distributions}
\label{sec:SupplPowerLaw}

\begin{figure}
	\centering
	\includegraphics[width=0.8\linewidth]{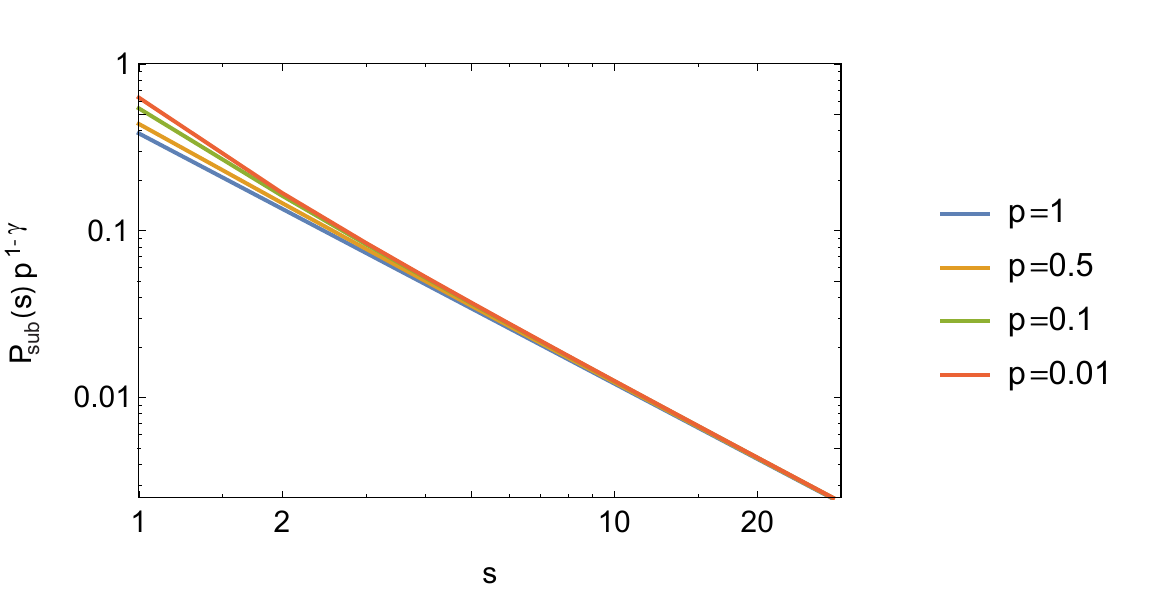}
	\caption{Scaling of subsampled power-law distributions $\Ps(s)$, using $a=1-\ga$, $b=0$, with $\ga=1.5$. the blue line shows the perfect power law of the fully sampled distribution, i.e. $P(s)$. Note the deviations from power law for small $s$, which increase with smaller sampling probability $\ps$.}
	\label{Fig:PLscal}
\end{figure}

To derive an approximate scaling for power-law distributions under subsampling, we expand on the work by Stumpf et al.~\cite{Stumpf2005}.  
Consider mathematical subsampling  as defined in the main text, and a power-law distribution $P(s)=C_\ga s^{-\ga}$ with exponent $\ga>1$, and normalization $C _\mathrm{\gamma}=1/\zeta(\gamma)$, {   where $\zeta(\gamma)$ is the Riemann zeta function}.
Then $\Ps(s;\ga,p)$ is a binomial subsampling with sampling probability $\ps$:  
\begin{equation}
\Ps(s;\ga,p)=C_\ga \sum_{n=0}^{\infty}(s+n)^{-\ga} p^s (1-p)^n {s+n \choose n}. \label{Eq:sub_prob}
\end{equation}
Building on the work by Stumpf et al., we assume that for the tail (i.e. for $s\rightarrow \infty$) the subsampled distribution is approaching an appropriately scaled power-law with slope $\gamma_\mathrm{sub}=\gamma$, i.e. we assume 
\begin{equation}
\Ps(s;\ga,p)\overset{s\to \infty}{\longrightarrow}c_{\gamma}(p)s^{-\ga}. 
\label{Eq:limit_sub}
\end{equation}
$c_\ga(p)$ is the subsampling-dependent normalization constant. To derive how $c_\ga(p)$ depends on $p$, we need to assume that
\begin{equation}
\frac{\partial}{\partial p}\Ps(s;\ga,p) 
\overset{s\to \infty} {\longrightarrow}s^{-\ga}\frac{\partial}{\partial p}c_\ga(p).
\label{Eq:ddpPsub}
\end{equation}
This is a strong assumption, because typically an exchange of differentiation and limit is only possible in case of uniform convergence of the derivatives~\cite{rudin1987real}, which is not the case here. However, all the functions we consider are monotonous in all parameters and numerical results support the assumption above.

In the following we assume that $s$ is large enough so that Eq.~\ref{Eq:limit_sub} can be taken as an identity. 
Then
\begin{eqnarray*}
	\frac{\partial}{\partial p}\Ps(s;\ga,p)
	&=&C_\ga s \sum_{n=0}^{\infty}(s+n)^{-\ga} p^{s-1} (1-p)^n {s+n \choose n}\\
	&-& C_\ga \sum_{n=0}^{\infty}(s+n)^{-\ga} p^s n (1-p)^{n-1} {s+n \choose n}.
\end{eqnarray*}
The first term can be approximated as:
\begin{eqnarray*}
	&&C_\ga s \sum_{n=0}^{\infty}(s+n)^{-\ga} p^{s-1} (1-p)^n {s+n \choose n} \\
	&=& \frac{s}{p} \Ps(s;\ga,p)
	 \approx \frac{s}{p}c_\ga(p)s^{-\ga}.
\end{eqnarray*}
The second term, after introducing $k=n-1$, reduces to:
\begin{eqnarray*}
	&& C_\ga \sum_{n=0}^{\infty}  (s+n)^{-\ga} p^s n (1-p)^{n-1} {s+n \choose n} \\
	& = & C_\ga \frac{s+1}{p}\sum_{k=0}^{\infty}(s+1+k)^{-\ga} p^{s+1} (1-p)^{k} {s+1+k \choose k}\\
	& = &\frac{s+1}{p} \Ps(s+1; \ga,p)	\\
	& \approx & \frac{s+1}{p} c_\ga(p)(s+1)^{-\ga}
\end{eqnarray*}

The ``$\approx$'' is inherited from Eq.~\ref{Eq:limit_sub}, which is only exact for $s \rightarrow \infty$.  Combining the two terms, we obtain for Eq.~\ref{Eq:ddpPsub}:

\begin{equation}
s^{-\ga}\frac{\partial}{\partial p}c_\ga(p)
\approx \frac{\partial}{\partial p}\Ps(s;\ga,p)
\approx \frac{s}{p}c_\ga(p)s^{-\ga}-\frac{s+1}{p} c_\ga(p)(s+1)^{-\ga}. 
\end{equation}

From this, $\frac{\partial}{\partial p}c_\ga(p)$ can be expressed as:

\begin{equation}
\frac{\partial}{\partial p}c_\ga(p)=\frac{c_\ga(p)}{p}\lim_{s\to \infty}\left[s-(s+1)\left(1+\frac{1}{s}\right)^{-\ga}\right]
\end{equation}

For solving the limit, we use the known identity

\begin{equation*}
\lim_{x \to 0}\frac{(1+x)^\mu -1}{\mu x}=1,
\end{equation*}

which can be restated by replacing $x$ by $1/s$ and $\mu$ by $-\ga$: 

\begin{equation*}
\lim_{s \to \infty}s\left(1+\frac{1}{s}\right)^{-\ga} -s=-\ga.
\end{equation*}

Thus

\begin{equation}
\frac{\partial}{\partial p}c_\ga(p)=\frac{c_\ga(p)}{p} (\ga-1) 
\end{equation}

This differential equation is solved by: 
\begin{equation}
c_\ga(p)=C^* p^{\ga-1},
\end{equation}
For $p=1$ we know that $C^*=C_\ga$, because sampling all units does not change the distribution. The final expression for $c_\ga(p)$ is thus
\begin{equation}
c_\ga(p)=C_\ga p^{\ga-1}. \label{Eq:C_ga}
\end{equation}
{   With this we can derive scaling parameters $a,b$ that collapse the distribution's tails, i.e. $\ps^a \Ps(\ps^b s)=P(s)$ for large $s$. Using Eq.~\ref{Eq:C_ga}:
\begin{equation}
\ps^a \Ps(\ps^b s) = C_\ga p^{\ga-1} \ps^a  (\ps^b s)^{-\gamma} = \ps^{a - b \ga +\ga-1} C_\ga s^{-\gamma}. 
\end{equation}	
	Thus for any $a$ and $b$, such that $a-b \gamma =1-\gamma$, the scaling ansatz leads to a collapse.	
	}
One of the members of this scaling family is $b=0$, $a=1-\ga$, which scales the $y$-axis only. As shown in Fig.~\ref{Fig:PLscal}, this scaling collapses the tails of distributions perfectly. For small $s$, however, there are systematic deviations under subsampling, which increase with smaller $\ps$. We call them ``hairs'', because they grow on the head of the distribution, as opposed to the tails. 

A different member of the scaling  family is $a=b=1$. This scaling is especially attractive, because it does not require information about the exponent $\gamma$ of the power law (see Fig.~\ref{Fig:MathSub}). As this scaling is linear in $p$, we call it p-scaling. 

{   \subsubsection{Power-law exponent close to unity} 
\label{sec:Exponent1}

Here we show why the ``hairs'' become smaller, i.e. converge to zero, in the limit of the power-law exponent $\gamma \to 1$. It is in agreement with results of Stumpf et al.~\cite{Stumpf2005}, stating that ``hairs'' are growing with increase of the exponent.   Mathematically, the exponent of the power-law distribution cannot be exactly equal to one or smaller, because in this case the distributions cannot be normalized. Thus without loss of generality we consider truncated power laws: $P(s) = C \cdot s^{-1}$ for $s\leq\smax$ and $P(s) = 0$ for $s>\smax$. The normalizing constant $C$ depends on  $\smax$. In this case the subsampled distribution $\Ps(s)$, with sampling probability $\ps$ can be written explicitly

\begin{equation*}
\Ps(s;\ps)= \sum_{l=s}^{\smax}\frac{C}{l}{l \choose s} p^s (1-p)^{l-s}=\frac{C}{s}\sum_{l=s}^{\smax}{l-1 \choose s-1} p^s (1-p)^{l-s}. 
\end{equation*}

We are interested in the behavior of the  ``hairs'' and thus consider small $s$. In this case, we can approximate $\Ps(s;\ps)$ by the infinite sum, and make use of the geometric series 

\begin{equation*}
\frac{1}{(1-x)^s}=\sum_{n=0}^{\infty} {n+s-1 \choose s-1} x^n
\end{equation*}

to obtain,  with a variable exchange $m=l-s$,

\begin{equation*}
\Ps(s;\ps)\approx \frac{C}{s} p^s \sum_{m=0}^{\infty}{m+s-1 \choose s-1}  (1-p)^{m}= \frac{C}{s}=P(s). 
\end{equation*}

Thus we showed that in the limit $\gamma \to 1$ subsampling of the power law converges to the original power law. 

}
\subsection{Inferring the system size from the subsampled distribution}
\label{sec:HairSize}
The deviations from power laws (i.e. the hairs), which emerge under subsampling, allow to infer the system size $M$ from the subsampled distribution $\Ps(s)$ alone, given that $P(s)$ follows a power law. 
This is because the hairs are a function of the sampling probability $\ps=N/M$. 
The hairs are  most pronounced for $\Ps(s=1)$ (except for $\Ps(s=0)$, which may remain unknown in experiments). {   Therefore, the inference of system size in experiments is most accurate if it is based on $\Ps(s=1)$. We explore this in the following derivations. Derivations based on other (small) $s$ can be performed analogously. }

Quantitatively, using the explicit relation for subsampling of power laws (Eq.~\ref{Eq:sub_prob}) with $l=n+1$ results in:

\begin{align*}
\Ps(s=1) &= \sum_{l=1}^{\infty} \frac{l^{-\gamma}}{\zeta(\gamma)} l (1-\ps)^{l-1} \ps \\
&= \frac{\ps}{(1-\ps) \zeta(\gamma)}\sum_{l=1}^{\infty} \frac{(1-\ps)^l}{l^{\gamma -1}} \\
&= \frac{\ps \cdot Li_{\gamma - 1}(1-\ps)}{(1-\ps)\zeta(\gamma)},
\end{align*}

where $\zeta$ is the Riemann zeta function, and $Li_{\gamma}(z)=\sum_{k=1}^{\infty}z^k/k^\gamma$ is the polylogarithm function. This relation is exact if $P(s)$ is a true power law.
For application to the real data obtained from subsampled observation the following algorithm allows to infer $\ps$:
\begin{enumerate}
	\item Check whether the experimentally obtained empirical distribution $P_{\mathrm{emp}}(s)$ is likely to originate from a system that under full  sampling shows a power-law distribution. If not, the method cannot be applied. 
	\item Estimate the power-law slope $\gamma$ of the power-law tail of the distribution to obtain $\hat{\gamma}$.
	\item Solve the following equation for $\ps$:
	
	\begin{equation*}
	P_{\mathrm{emp}}(s=1)= \frac{\ps \cdot Li_{\hat{\gamma} - 1}(1-\ps)}{(1-\ps)\zeta(\hat{\gamma})}
	\end{equation*}
	
\end{enumerate}
This will return the sampling probability $\ps$. From this, the system size can be inferred if $N$ is known. 
This approach is also applicable approximately if the full system does not display a pure power law, but a power law with cutoff at large $s$. Then the power-law slope $\gamma$ has to be inferred on an appropriate interval between the hairs and the cutoff.

\begin{figure}
	\centering
	\includegraphics[width=0.49\linewidth]{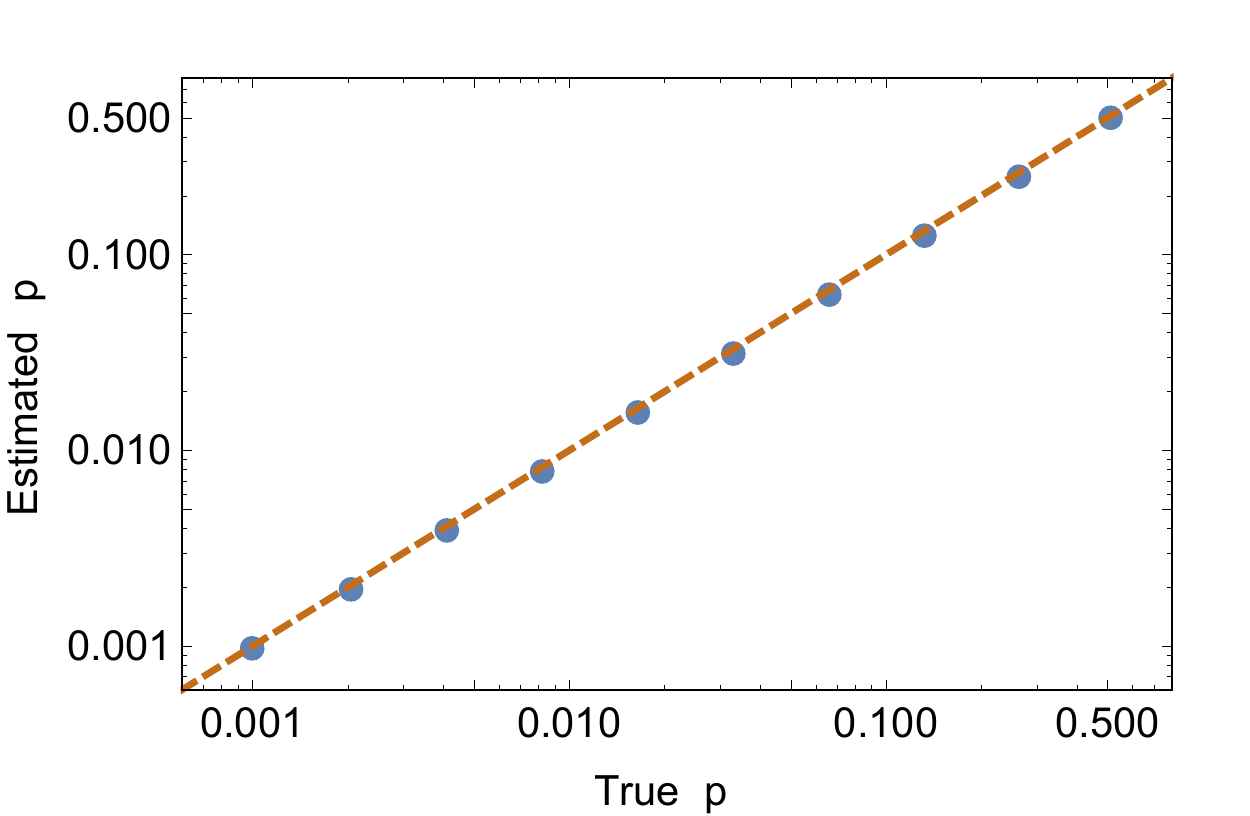}
	\caption{Estimating the sampling fraction $\ps$ from the ``hairs''. Dots represent the estimated $\ps$ as a function of the true one, the dashed line represent the perfect correspondence.}
	\label{Fig:HairSize}
\end{figure}

We applied this method numerically to the data generated by the critical branching model of size $M=1024$, subsampled to $N=2^0, 2^1,\ldots 2^9$ units based on $10^7$ avalanches in the full system. 
Indeed, the full system size could be inferred by $\hat{M}=\hat{\ps} N$ with high precision (Fig~\ref{Fig:HairSize}): The maximal deviations were smaller than 6\%.

\subsection{Subcritical systems}
\label{Suppl:Subcr}

\begin{figure}
	\centering
	\includegraphics[width=0.49\linewidth]{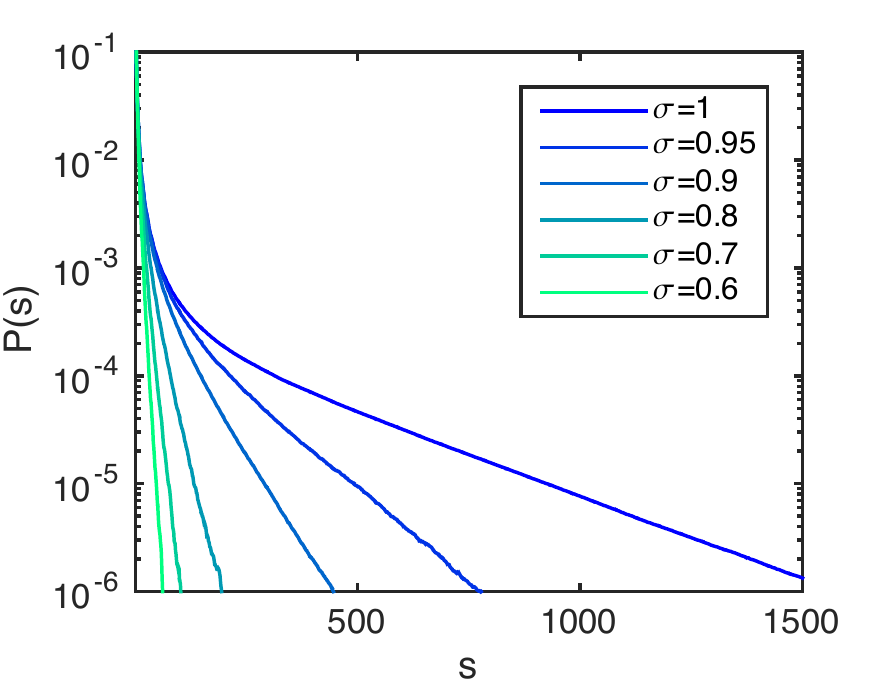}
	\includegraphics[width=0.49\linewidth]{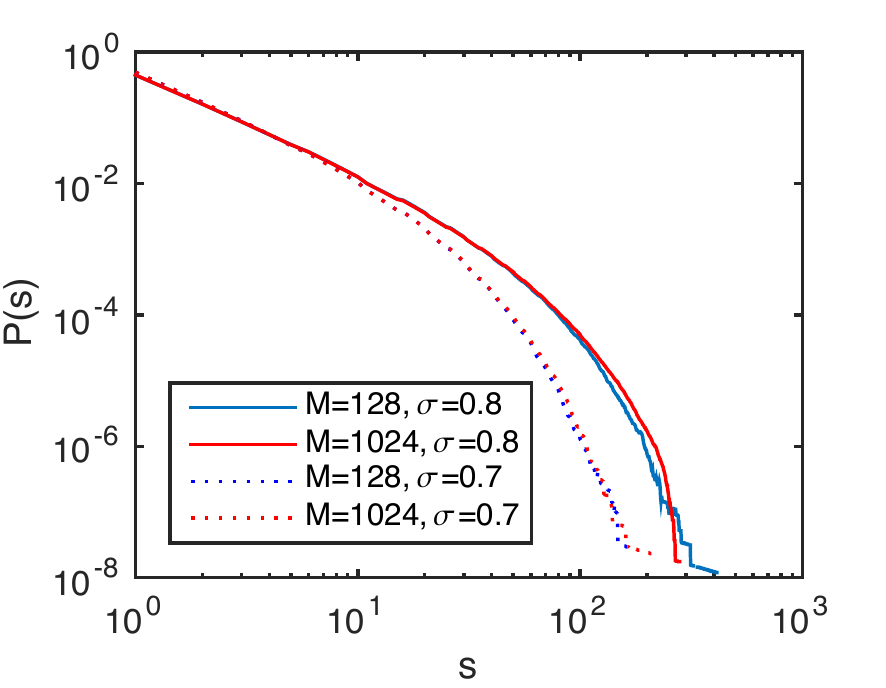}
	\caption{Exponential tails of subcritical distributions.  Left: Subcritical distributions for different branching ratios $\sigma$ plotted in a log-lin scale clearly show exponential tails, with the tail slope $\lambda$ depending on $\sigma$ (results for $M=1024$). Right: Subcritical distribution with a fixed deviation from criticality ($\sigma=0.7$ or 0.8) for different system sizes $M$.} 
\label{Fig:SubExpTail}
\end{figure}

As outlined in the main text, avalanche distributions collapse under p-scaling for critical systems, but not for subcritical systems.
The main reason is that for subcritical systems the exponential tail is too steep, i.e. the requirement  $\lambda \ll \ps$ is violated. We in the following derive an approximate relation between $\lambda$ and the distance to criticality ($\varepsilon = \sigma_\mathrm{crit}-\sigma= 1-\sigma$). We show that for more subcritical systems, $\lambda$ becomes increasingly larger (see also Fig.~\ref{Fig:SubExpTail} (left)).   
To approximately derive the relation between $\lambda$ and $\varepsilon$ (or $\sigma$), we used the branching process~\cite{Harris1963}, because it allows to easily control the distance to criticality by changing the branching ratio $\sigma$, and because it is independent of finite size effects. This is a reasonable assumption, because in subcritical systems $P(s)$ is not affected by changing the system size for any $M>M_0$. Only for very small systems sizes there are finite size effects (Fig.~\ref{Fig:SubExpTail} (right)).

\begin{figure}
	\centering
  \includegraphics[width=0.49\linewidth]{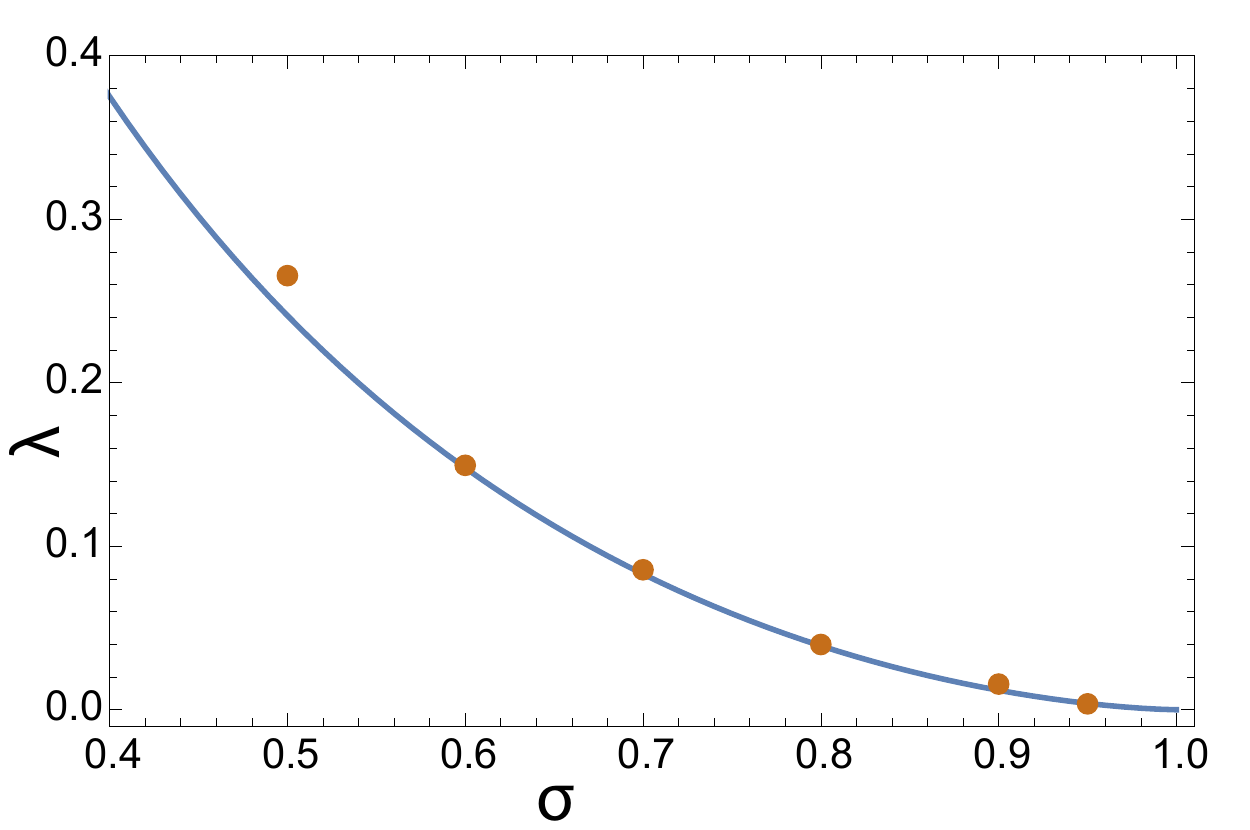}
	\caption{Slope of tails, $\lambda$, for avalanche distributions of subcritical models.
	$\lambda$ increases with increasing distance to criticality (decreasing $\sigma$). The dots denote the numerical results for $10^7$ avalanches on the full system, the line denotes the analytical results.} 
	\label{Fig:SigmaLamda}
\end{figure}

To derive heuristically the slope of the exponential tail $\lambda$ as a function of the control parameter $\sigma$, consider a  branching process with branching ratio $\sigma<1$, and assume that an avalanche starts with 1 neuron firing. Then on expectation in the second time step there are $\sigma$ neurons firing, in the third time step $\sigma^2$, and so forth. Thus we obtain an expression for the average avalanche size $\langle s \rangle$:

\begin{equation*}
\langle s \rangle = 1+\sigma+\sigma^2+\sigma^3+\ldots = \frac{1}{1-\sigma}.
\end{equation*}

In the subcritical regime, the distribution of the avalanche sizes is dominated by the exponential cutoff. We consider that they are well approximated by the power law with slope $\gamma$ and an exponential cutoff parametrized by $\lambda$

\begin{equation*}
P(s)=C_{\mathrm{norm}} s^{-\ga} e^{-\lambda s}.
\end{equation*}

The mean value of this distribution is given by:

\begin{equation*}
\langle s \rangle = \frac{Li_{1-\ga}(e^{-\lambda})}{Li_{\ga}(e^{-\lambda})}
\end{equation*}

Where $Li_{\gamma}(z)=\sum_{k=1}^{\infty}z^k/k^\gamma$ is again the polylogarithm function.
The relation between $\lambda$ and $\sigma$ is thus

\begin{equation}
\frac{Li_{1-\ga}(e^{-\lambda})}{Li_{\ga}(e^{-\lambda})}=\frac{1}{1-\sigma}, 
\label{eq:lamsig}
\end{equation}

and hence $\lambda$ approaches zero when approaching the critical point ($\sigma \rightarrow 1$). 
As $\lambda$ decays slowly as a function of $\sigma$, except in the very close vicinity of the critical point,
the requirement for p-scaling, $\lambda \ll \ps$, is only satisfied in the close vicinity of the critical point. Else p-scaling does not apply.

To compare our analytical with numerical results, we used the same data as in Fig.~\ref{Fig:SubExpTail}. We first estimated $\gamma \approx 1.3$ from the distributions, and with this solved equation~\ref{eq:lamsig} numerically. 
The analytical results closely fitted the slopes $\lambda$ of the exponentials from the simulations (Fig~\ref{Fig:SigmaLamda}).


{  
\subsection{Subsampling scaling for the EHE-model and the sparsely connected branching model}
\label{sec:EHE_sparseBM}

In this section, we investigate whether subsampling scaling also applies to other models than the ones treated in the main manuscript. In particular, we treat here first the Eurich, Herrmann \& Ernst (EHE) model~\cite{Eurich2002}, a classical extension of the BTW model to neural networks, and then a realization of the BM with sparse connectivity ($k=4$, see Methods). The details of the EHE model can be found in~\cite{Eurich2002,LevinaDiss}. This model produces power-law distributions of the avalanche sizes with slope $\approx 1.5$ that indicates that it belongs to the same university class as the branching model. 
 However, activity transmission is not stochastic as in BM, but deterministic as in BTW.   Another peculiarity of the model lays in its dissipative nature: for the finite system sizes $M$ each spike leads to dissipation of $\Delta\approx 1/\sqrt{M}$~\cite{Eurich2002}. Thus only in the limit $M \to \infty$ the model is both truly critical and conservative.  We simulated the EHE model with both, the classical fully connected graph topology and also with random connectivity probability $\pconn$. For both models, the dynamics is as follows: Each neuron $i$ is a non-leaky integrator, and its membrane potential is denoted by $h_i\in[0,1)$. When $h_i$ crosses the threshold $\theta=1$ the neuron fires and is reset $h_i \mapsto h_i-1$ and all its postsynaptic connections receive an input of strength $\alpha / M$, where $\alpha$ is the control parameter in the model, the strength of interaction. For the fully connected network, it is known that $\alpha\approx 1-M^{-0.5}$ leads to an approximate power-law distribution of the avalanche sizes. For not-fully connected networks the connection probability $\pconn$ needs to be included, and thus the condition to achieve approximate power-law distributions generalizes to $\alpha \cdot \pconn\approx 1-M^{-0.5}$.
 
 As the avalanche size distribution in the EHE model can be directly mapped to the branching model~\cite{LevinaDiss}, subsampling scaling is expected to behave the same as in the BM, producing ``hairs'' but resulting in a good collapse. We tested this for the model of $M=1024$ neurons with $\pconn=0.1$ and obtained, as expected, a collapse under subsampling scaling (Fig.~\ref{Fig:EHE_sub}).

The distributions of the fully and the sparsely connected BM are very similar (Fig. \ref{Fig:ModelSubSuppl}). The only difference is a slightly more pronounced lack of small avalanches in the fully sampled sparse BM (Fig. \ref{Fig:ModelSubSuppl}~C), which translates to somewhat less pronounced ``hairs'', in particular under ``mild'' subsampling ($N\geq 2^{10}$).

 \begin{figure}
 	\centering
 	\includegraphics[width=.75\linewidth]{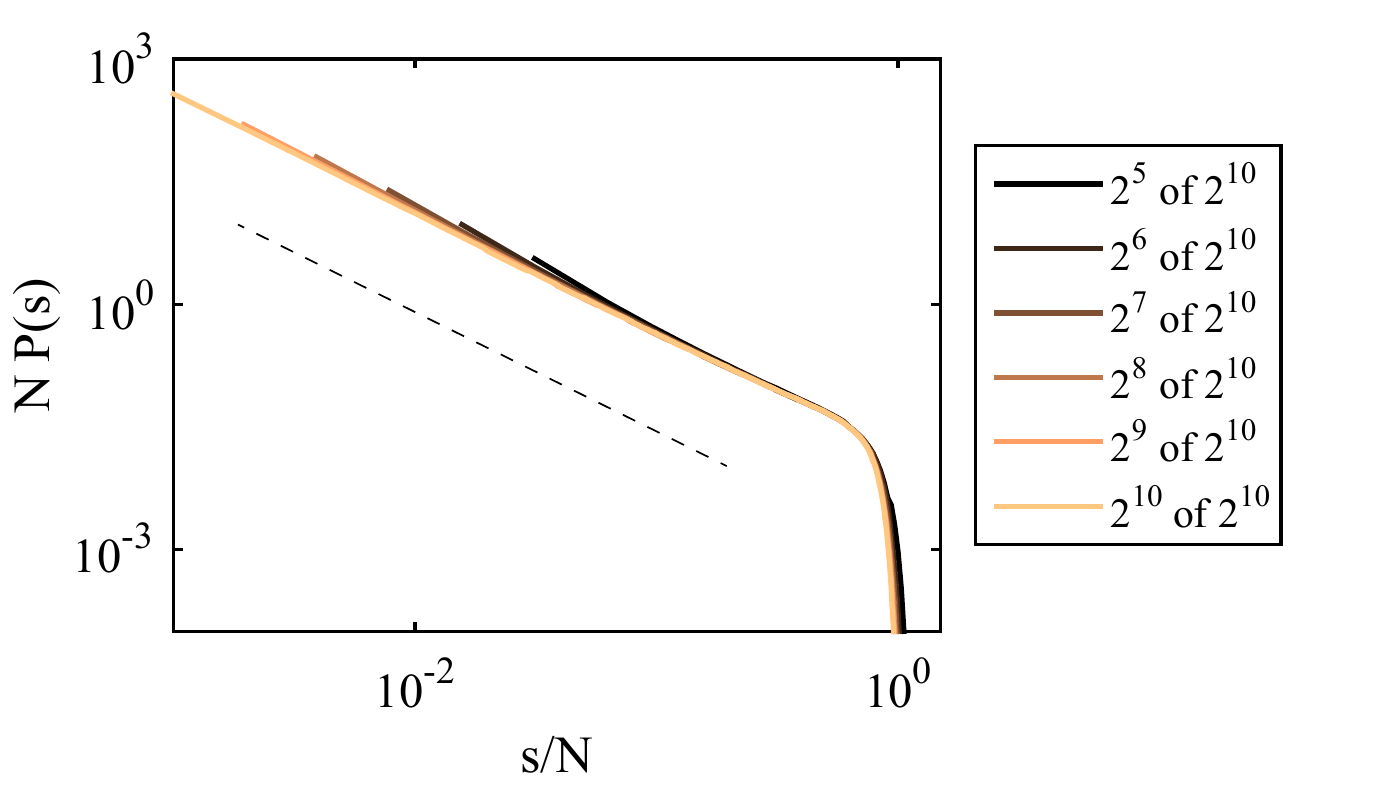}
 	\caption{{  
 		Subsampling scaling of the EHE model. $N=32, 64,\dots, 512$ units were sampled out of $M=1024$ units. Parameters: connection probability $\pconn=0.1$, connection strength $\alpha=0.96/(M\cdot \pconn)$. The dashed line indicates a slope of $-1.5$.}}
 	\label{Fig:EHE_sub}
 \end{figure}

	\begin{figure}
	\centering
	\includegraphics[width=.89\linewidth]{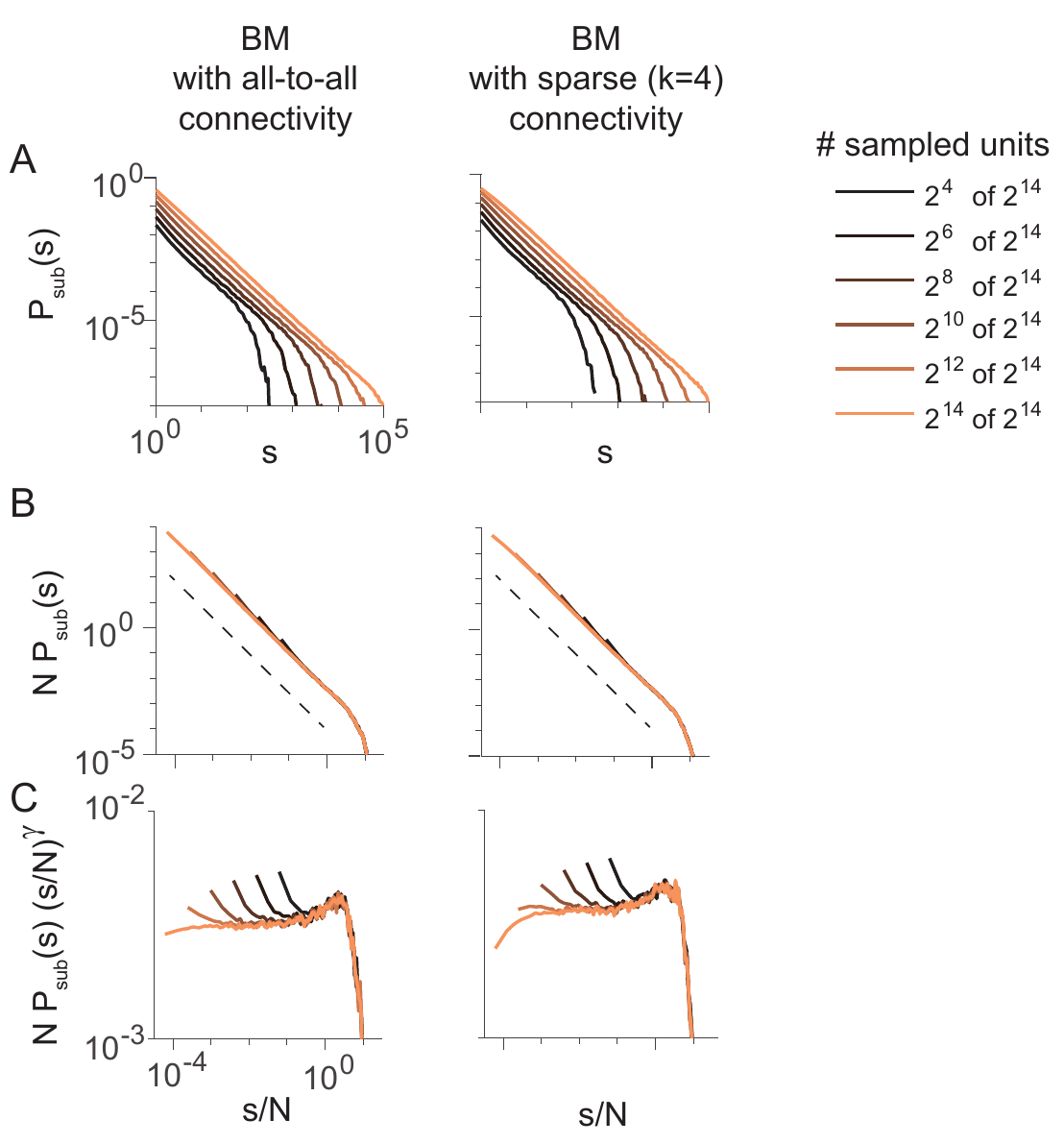}
	\caption{   Subsampling scaling in the fully and the sparsely connected branching model (BM), corresponding to Fig. \ref{Fig:ModelSub} in the main text. Here avalanche size distributions $\Ps(s)$ are shown for the BM with all-to-all connectivity (left), as in the main text, and for the BM with sparse, annealed connectivity with degree $k=4$ (right). 
	\textbf{A}: Both versions of the BM show very similar distributions, 
	\textbf{B}: and a good collapse. 
	\textbf{C}: In the flattened representation, minor differences between the two models become apparent, namely a lack of small avalanches $s$ in the fully sampled, sparsely connected version of the BM.
	Both variants of the model have slope $\ga=1.5$ (dashed black line).}
	\label{Fig:ModelSubSuppl}
\end{figure}

}

\vspace{0.5cm}

{   
\subsection{Detailed discussion of the experimental results}\label{Sec:SI_Experiments}

\begin{figure}
	\centering
	\includegraphics[width=.89\linewidth]{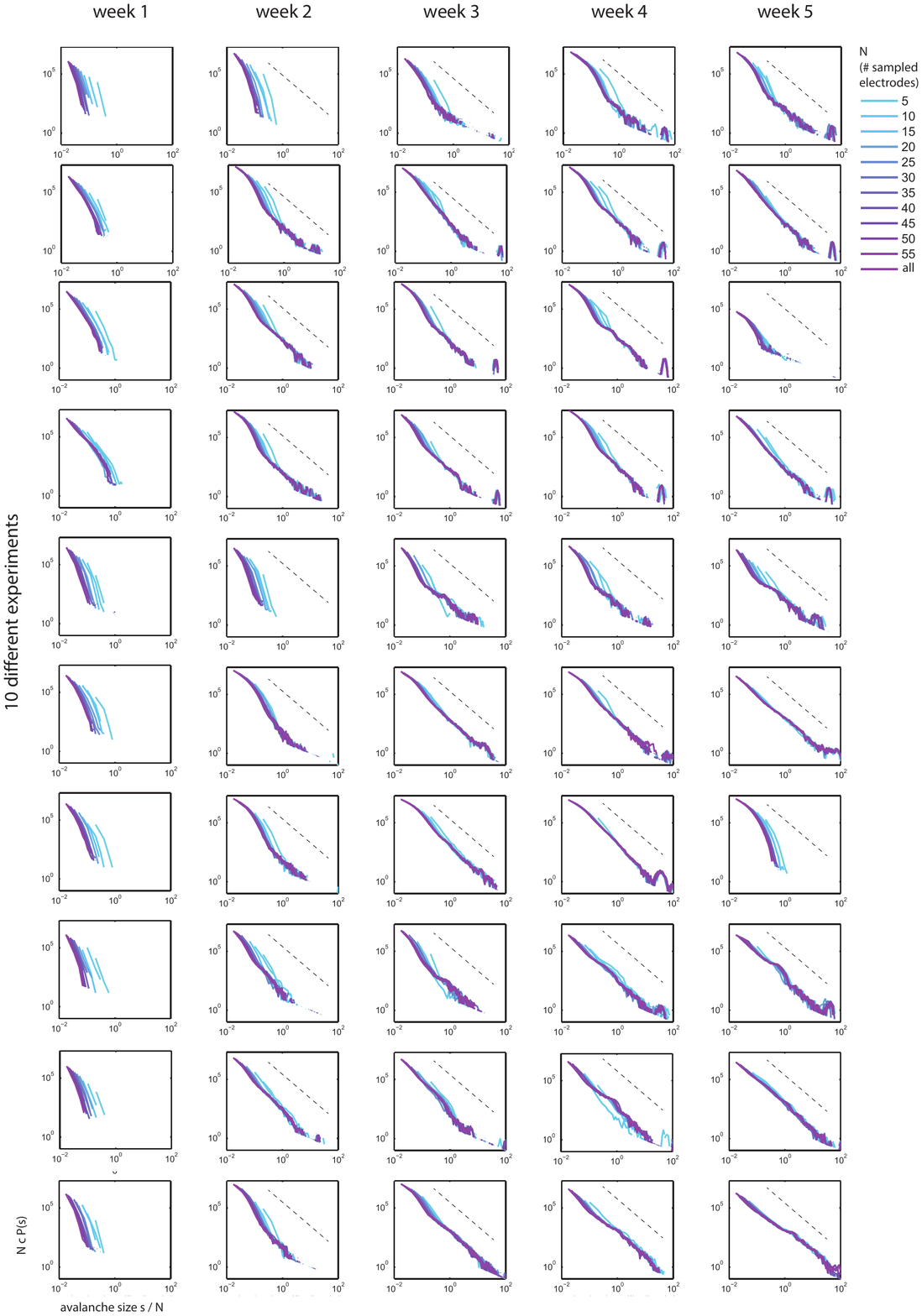}
	\caption{Changes of the avalanche size distributions with development. This figure corresponds to Fig.~\ref{fig:PLscal} in the main text, but here shows distributions for all recordings we evaluated, and for all five recording weeks (typically day 7, 14, 21, 28, 34). For each experiment, the p-scaled avalanche size distributions $\Ps(s)$ are displayed; $c$ denotes the total number of avalanches observed in the respective recording, and the dashed line a slope of $-2$ for visual guidance.}
	
	\label{Fig:SpikeFSall}
\end{figure}

Figure~\ref{Fig:SpikeFSall} displays $\Ps(s)$ for all recordings of developing neural cultures we evaluated (see Methods). As discussed in the main text (Fig.~\ref{fig:PLscal}), with maturation the $\Ps(s)$ approached power-law scaling, which for the fully sampled culture is expected to extend over almost six orders of magnitude. In addition to the power laws,  about half of the mature cultures also showed a bump in $\Ps(s)$ at very large sizes ($s \approx 5000$). These very large avalanches comprise only a tiny fraction of all avalanches ($\approx 0.02 \%$). Such bumps are a priori not expected for critical systems. The collapse of the bumps itself is a manifestation of the activity spread during the large avalanches that hit the sampled set proportionally to the number of sampled units. In the following we discuss first whether the distributions with the bumps are expected to collapse under p-scaling, and then the potential origin of the bumps.

Regarding the questions whether the distributions observed here are expected to collapse, the answer is straight forward: The avalanches in the tail make only a tiny fraction of all observed avalanches (about 2 in 10,000), while the other 99.98\% avalanches follow a power law for about 3 orders of magnitude. (It is the log-log scale together with the logarithmic binning that might make the bumps appear more prominent than they are.) With only 0.02\% of avalanches not following a power law, a decent collapse is to be expected.  

Regarding the origin of the bumps, there are a number of potential explanations, which we outline in the following. At first glance, the bumps are reminiscent of supercritical systems (hypothesis 3), however, they do not occur at $N$, the number of sampled units, as expected for supercritical systems. More likely, they may represent finite size effects (hypothesis 1), or alternatively transient switches to a bursty state (hypothesis 2). All three hypotheses are detailed in the following:

\begin{enumerate}
\item 	\textit{Biological finite size effects in a critical system.} \\
Assume the neural cultures were precisely at a critical point. Thus the distribution of the avalanche sizes would be a perfect power law without cutoff. 
However, in biological systems the avalanche size cannot go to infinity,   because biological mechanisms (e.g. depletion of synaptic resources, shortage of Ca$^{2+}$ or  homeostatic mechanisms) limit their maximal size. 
All these avalanches that are larger than some $\str$ (in our data $\str \approx 3000$) are thus expected to be distributed around a characteristic, biologically determined size, which here is about $s \approx 5000$. 
The probability $\pb$ to observe an avalanche larger than some maximal size $\str$ is given by the Hurwitz zeta function. 
Indeed, in agreement with this hypothesis, the number of the avalanches observed in the “bump” agrees with the probability $\pb$ for perfectly critical system. Thus the data support our hypothesis that the bump represents the collection of all avalanches that would, in an ideal system, be larger than 3000. (Note that all avalanche sizes $s$ given here are the sizes observed under subsampling).
Thus biological finite size effects are a probable origin for the bumps.

\item \textit{Criticality alternates with a state that gives rise to large avalanches} \\
The \textit{in vitro} neural networks we analyzed could in principle alternate between different states. While in one state, which comprises about 99.98\% of the avalanches, the system is critical, in the other state it displays unusually large avalanches that run multiple times over the entire system and give rise to population bursts, i.e. they manifest as the observed bumps. The precise fraction of ``burst avalanche'' can depend on the properties of each individual culture (some showing none at all), and it could be pure coincidence that the fraction of burst avalanches is in agreement with the fraction expected for the avalanche tail (see hypothesis 1).

\item	\textit{A novel form of slight supercriticality in a finite system.} \\
While it is straight forward to identify ``\textit{sub}criticality'' (no avalanches covering the full system size, no power-law behavior of distributions, but a prominent exponential tail), it is much trickier to identify ``supercriticality'' in neural systems by pure observation, potentially because supercriticality in the thermodynamic limit implies a non-zero fraction of infinite avalanches, but in finite systems it depends on the type of system how these infinite avalanches manifest.
For supercritical systems in neuroscience, the bump in $\Ps(s)$ occurs typically at $N$, i.e. the system size or the number of sampled neurons \cite{Beggs2003, Levina2007}. However, here in all experiments where the bump is observed, it is around 80 times $N$ (i.e. $s \approx 5000$ from sampling up to 60 electrodes). Thus here the bumps do not indicate supercritical behavior resembling that of previous studies. 
However, it could indicate a novel form of supercriticality on a finite system.
\end{enumerate}

How to distinguish between these potential causes of the bump appearance remains an open question for further experimental investigations (e.g. changing the network size; making the network on purpose supercritical).
In the experiments evaluated here, the presence of the data collapse in the more mature networks predicts a power-law distribution for $P(s)$ of the full neural system that spans approximately 6 orders of magnitude. However, whether such power-laws scaling is sufficient to infer criticality, is still under debate. 

}

{  \subsection{Combining subsampling scaling and finite-size scaling}

\label{Sec:FSS_sub}

	As demonstrated in section~\ref{S:sub_vs_fss}, there is a fundamental difference between subsampling scaling that deals with partial observations of a system, and finite-size scaling (FSS) that extrapolates from models of finite size to infinite size systems. Here we show how to combine both scaling ans\"atze to obtain a universal scaling. 
	
	The finite-size scaling ansatz for a critical system is formulated as:
	\begin{equation}
	P(s,M)=M^{-\beta}g\left(\frac{s}{M^\nu}\right)\: \Leftrightarrow\: M^\beta P(s M^\nu,M) = g(s), \label{Eq:FSS}
	\end{equation}
	where $g(s)$ is a scaling function. 
	The formulation for the subsampling scaling in a system with $M$ units is:
	
	\begin{equation*}
	\Ps (s,N;M)=N^{-1} \gs\left(\frac{s}{N};M\right). 
	\end{equation*}
	
	This can be re-written as:
	
	\begin{equation}
	N \Ps(s N,N;M)=\gs(s;M)=M \Ps(s M,M;M). \label{Eq:SubsScalasFSS}
	\end{equation}
	
	Our goal is to combine the finite size scaling and the subsampling scaling relations (Eqs.~\ref{Eq:FSS} and \ref{Eq:SubsScalasFSS}) to factorize out the dependence of $\gs$ on $M$, and hence be able to collapse subsampled distributions from different system sizes $M$. 
	To find the appropriate scaling, we need to identify the exponents $\delta$ and $\kappa$ such that 
	
	\begin{equation*}
	N M^\delta \Ps(s N M^\kappa,N;M)=g(s).
	\end{equation*}
	
	To this end, recall that $\Ps(s,M;M)=P(s,M)$. In the following we first use subsampling scaling to express $\Ps(s,M;M)$ in terms of $P(s,M)$ using Eq.~\ref{Eq:SubsScalasFSS}:
	
	\begin{eqnarray*}
		N M^\delta \Ps(s N M^\kappa,N;M)=M^{1+\delta} \Ps(s M^{\kappa+1},M;M)= \\
		M^{1+\delta} P(s M^{\kappa+1},M) = M^{1+\delta-\beta} g(s M^{\kappa+1-\nu}).
	\end{eqnarray*}
	
	Thus the solutions for the exponents is given by $\delta=\beta-1$ and $\kappa=\nu-1$, and hence the general subsampling-finite-size scaling is given by:
	
	\begin{equation}
	N M^{\beta-1} \Ps(s N M^{\nu-1},N;M)=g(s). \label{Eq:FSS_sub}
	\end{equation}
	
	We tested this relation numerically for the case of the branching model (BM). For this model, FSS is given by $\beta=1.5$, $\nu=1$ and thus the subsampling-finite-size scaling is given by:	
	
	\begin{equation}
	P(s)=N M^{0.5} \Ps(s N). \label{Eq:Scal_comb}
	\end{equation}
	
	Indeed, with this scaling we obtained as expected a good collapse for combining different sampling sizes $N$ and system sizes $M$ (Fig.~\ref{Fig:FSS_subs}).

}

{   \subsection{Numerical estimation of optimal scaling} \label{Sec:NumApproach}
\begin{figure}
	\centering
	\includegraphics[width=.95\linewidth]{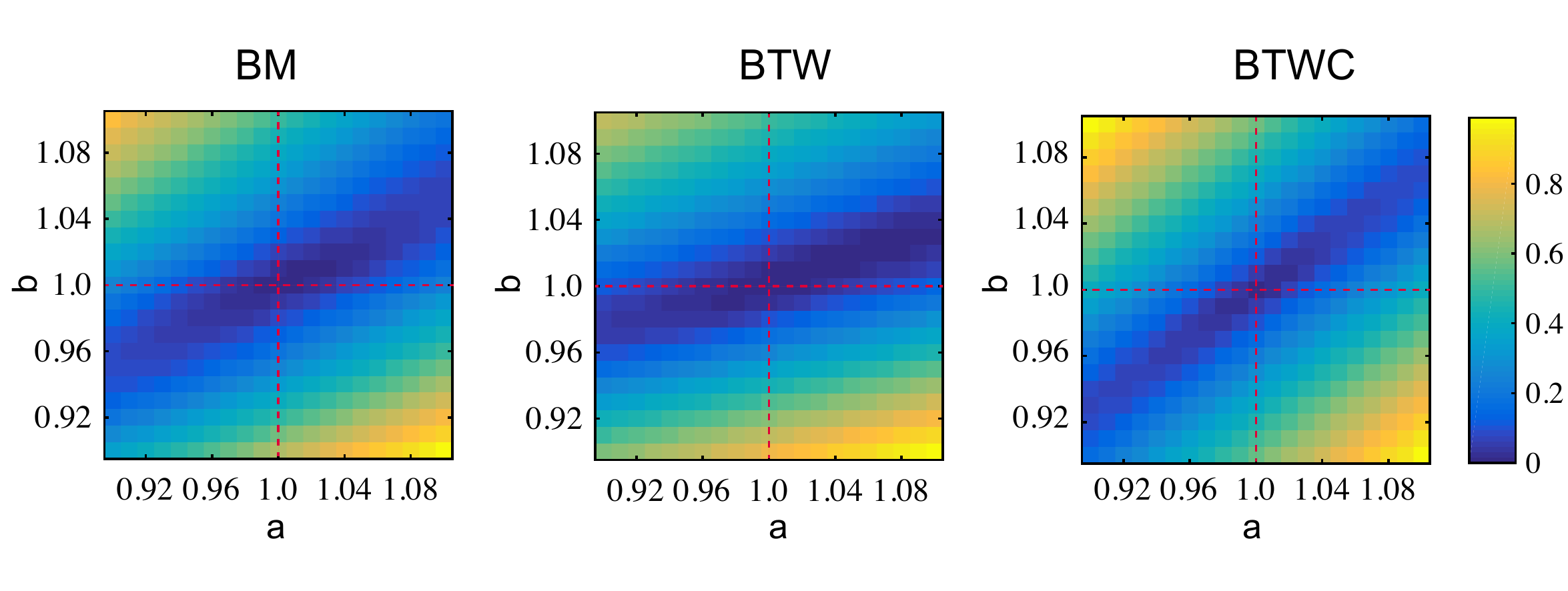}
	\caption{   Numerical estimation of scaling parameters $a,b$. Color code represent deviation from the perfect collapse $d(a,b)$, yellow -- large deviation,  dark blue -- close to perfect collapse. Red dashed lines denote the analytical prediction. Left: Branching model (BM), middle: Bak-Tang-Wiesenfeld model (BTW), right: BTW with circular boundaries (BTWC). For presentation purposes $d(a,b)$ was shifted and scaled to the interval between zero (for the minimal value) and unity. This procedure does not change the location of the minima.}
	\label{Fig:NumScal}
\end{figure}

Throughout the manuscript we used an analytical approach to determine the optimal scaling for the subsampled distributions. In this section we confirm that our analytical results coincide with a direct numerical estimation of the scaling constants $a,b$. 
To achieve an optimal scaling collapse, we numerically estimated the parameters $a$ and $b$ that minimize the distance $d(a,b)$ between the rescaled distributions $\Ps(s;a,b,N)=N^a \Ps(N^b s)$ under subsampling, and the rescaled distribution under full sampling, $P(s;a,b,N=M)$. In more detail, we first estimated for each $N \in [8,16,32, \dots , M]$ the distance $d(a,b;N)$ as follows:
\begin{equation}
d(a,b,N) = \left\langle|\ln(\Ps(s;a,b,N)) - \ln(P(s';a,b,M))|\right\rangle_s.
\end{equation}
The mean $\left\langle \cdot\right\rangle_s$ was taken over all $s \in [1, 10\cdot N]$, and $s'$ are the support points in $P(s';a,b,M)$ corresponding to those in $\Ps(s;a,b,N)$, i.e. $s' / M^b = s / N^b$ is fulfilled. As $s'$ may take non-integer values, the values $\Ps(s';a,b,N)$ are obtained, if necessary, by linear interpolation between the nearest integers.
Then the weighted average over all $d(a,b,N)$ is taken to obtain $d(a,b)=\left\langle d(a,b,N)\right\rangle_N$. The parameter combination $(a^*, b^*)$ that minimizes $d(a,b)$ provides numerically the optimal collapse.  We scanned $a$ and $b$ in steps of $0.01$ and found for both, the BM and the BTWC an optimal collapse at $a^*=b^*=1$, as predicted analytically (Fig.~\ref{Fig:NumScal}). For the BTW, the optimal collapse was at $a^*=1.01, b^*=1.02$, but the value of $d(a,b)$ in the point analytically obtained $a=b=1$ deviated only by $3\%$ from the absolute minimum  (Fig.~\ref{Fig:NumScal}).
Thus overall, our numerical results match very well the theoretical prediction.

}